\documentclass[a4paper,11pt]{article}
\pdfoutput=1 

\usepackage{jheppub} 
\usepackage{soul}

\usepackage[T1]{fontenc} 

\usepackage{placeins}

\usepackage{blindtext,hyperref}
\usepackage{amsfonts,amsmath,amscd}
\usepackage[format=hang,singlelinecheck=0,font={sf,small},labelfont=bf,caption=false]{subfig}
\usepackage{cleveref}
\Crefname{equation}{Eq.}{Eqs.}
\Crefname{figure}{Fig.}{Figs.}
\Crefname{tabular}{Tab.}{Tabs.}
\Crefname{section}{Sec.}{Secs.}

\graphicspath{{figures/}}

\title{\boldmath Metastable attractors explain the variable timing of stable behavioral action sequences}

\author[1,a,b]{Stefano Recanatesi\note{Co-first authors.}}
\author[1,c,d]{Ulises Pereira}
\author[e,f]{Masayoshi Murakami}
\author[2,f]{Zachary Mainen\note{Co-corresponding authors.}}
\author[2,b,g]{Luca Mazzucato}
\affiliation[a]{University of Washington, Center for Computational Neuroscience and Swartz Center, Seattle}
\affiliation[b]{Institute of Neuroscience, University of Oregon, Eugene.}
\affiliation[c]{Department of Statistics, The University of Chicago, Chicago}
\affiliation[d]{Center for Neural Science, New York University, New York}
\affiliation[e]{Department of Neurophysiology, University of Yamanashi, Japan.}
\affiliation[f]{Champalimaud Centre for the Unknown, Lisbon, Portugal.}
\affiliation[g]{Departments of Biology and Mathematics, University of Oregon, Eugene.}

\emailAdd{zmainen at neuro dot fchampalimaud dot org; lmazzuca at uoregon dot edu}

\abstract{Natural animal behavior displays rich lexical and temporal dynamics, even in a stable environment. This implies that behavioral variability arises from sources within the brain, but the origin and mechanics of these processes remain largely unknown. Here, we focus on the observation that the timing of self-initiated actions shows large variability even when they are executed in stable, well-learned sequences. Could this mix of reliability and stochasticity arise within the same circuit? We trained rats to perform a stereotyped sequence of self-initiated actions and recorded neural ensemble activity in secondary motor cortex (M2), which is known to reflect trial-by-trial action timing fluctuations. Using hidden Markov models we established a robust and accurate dictionary between ensemble activity patterns and actions. We then showed that metastable attractors, representing activity patterns with the requisite combination of reliable sequential structure and high transition timing variability, could be produced by reciprocally coupling a high dimensional recurrent network and a low dimensional feedforward one. Transitions between attractors were generated by correlated variability arising from the feedback loop between the two networks. This mechanism predicted a specific structure of low-dimensional noise correlations that were empirically verified in M2 ensemble dynamics. This work suggests a robust network motif as a novel mechanism to support critical aspects of animal behavior and establishes a framework for investigating its circuit origins via correlated variability.
}

\begin{document} 
\maketitle
\flushbottom

\section{Introduction}
\label{sec:intro}

When interacting with a complex environment, animals generate naturalistic behavior in the form of self-initiated action sequences, originating from the interplay between external cues and the internal dynamics of the animal. Self-initiated behavior exhibits variability both in its temporal dimension (when to act) and in its spatial features (which actions to choose, in which order) \cite{berman2016predictability,wiltschko2015mapping,markowitz2018striatum}. Large trial-to-trial variability has been observed in action timing, where transitions between consecutive actions are well described by a Poisson process \cite{killeen1988behavioral}. Recent studies in C. Elegans \cite{linderman2019hierarchical}, Drosophila \cite{berman2016predictability} and rodents \cite{wiltschko2015mapping,markowitz2018striatum} demonstrated that the spatiotemporal dynamics of self-initiated action sequences can be captured by state space models, based on an underlying Markov process. These analyses revealed a repertoire of behavioral motifs typically numbering in the hundreds, leading to a combinatorial explosion in the number of action sequences. Such a large behavioral landscape poses a formidable challenge for investigating the neural underpinnings of behavioral variability. A promising approach to tame the curse of dimensionality is to reduce the lexical variability in the behavioral repertoire, by using a task where the set of actions is rewarded when executed in a fixed order, yet retaining variability in action timing \cite{murakami2014neural,murakami2017distinct}, a hallmark of self-initiated behavior \cite{killeen1988behavioral}.

Previous studies in rodents have identified the secondary motor cortex (M2) as part of a distributed network involved in motor planning, working memory \cite{li2016robust} and self-initiated tasks \cite{murakami2014neural,murakami2017distinct}. During delay periods in decision-making tasks, trial-averaged population activity in M2 displays clear features of attractor dynamics, with two discrete attractors encoding the animal's upcoming choice \cite{inagaki2019discrete}. Are attractor dynamics in M2 specific to delay period activity? Here, we investigate the alternative hypothesis that attractor dynamics represent the intrinsic operational regime of M2 neural circuits throughout sequences of self-initiated behavior. In particular, we sought to uncover a correspondence between M2 attractors and upcoming self-initiated actions.

Because self-initiated action sequences are characterized by large trial-to-trial temporal variability in transition timing, they cannot be directly aligned across trials, hampering the applicability of traditional trial-averaged measures of neural activity. A principled framework to tackle this issue is to model neural population dynamics using hidden Markov models (HMMs) \cite{Rabiner1989}. These state space models can identify hidden states from population activity patterns in single trials, and have been successfully deployed in a variety of tasks and species from C. Elegans \cite{linderman2019hierarchical} to rodents \cite{Jones2007,mazzucato2015dynamics,maboudi2018uncovering,la2019cortical}, primates \cite{gat1993statistical,Abeles1995a,PonceAlvarez2012,engel2016selective} and humans \cite{baldassano2017discovering,taghia2018uncovering}. Hidden Markov models segment single-trial population activity into sequences in an unsupervised manner by inferring hidden states from multi-neuron firing patterns. Within each pattern, neurons fire at an approximately constant firing rate for intervals typically lasting hundreds of milliseconds. 

Previous work showed that the activity patterns, revealed by hidden Markov models, can be interpreted as metastable attractors, arising from recurrent dynamics in local cortical circuits \cite{MillerKatz2010,mazzucato2015dynamics}. Metastable attractors are produced in biologically plausible network models \cite{DecoHugues2012,LitwinKumarDoiron2012} and have been used to elucidate features of sensory processing \cite{MillerKatz2010,mazzucato2015dynamics}, working memory \cite{Amit1997b} and expectation \cite{mazzucato2019expectation}, and to explain state-dependent modulations of neural variability \cite{DecoHugues2012,LitwinKumarDoiron2012,mazzucato2016stimuli}. However, while previous models are capable of generating sequential activity \cite{sompolinsky1986temporal,kleinfeld1986sequential,MillerKatz2010,pereira2018unsupervised}, they are hindered by a fundamental trade-off between sequence reproducibility and trial-to-trial temporal variability. Namely, they can endogenously generate either reliable sequences without temporal variability \cite{sompolinsky1986temporal,kleinfeld1986sequential,pereira2018unsupervised} or, instead, sequences with large temporal variability but unreliable order \cite{LitwinKumarDoiron2012,mazzucato2015dynamics}. Thus, existing models are incapable of generating reproducible sequences of metastable attractors, characterized by large trial-to-trial variability in attractor dwell times.

Here, we addressed these issues in a waiting task \cite{murakami2014neural,murakami2017distinct} in which freely moving rats performed many repetitions of a sequence of self-initiated actions leading to a water reward. The identity and order of actions in the sequence was fixed by the task reward contingencies (i.e., producing out-of-sequence actions yielded no rewards), yet action timing retained large trial-to-trial variability \cite{murakami2014neural,murakami2017distinct}. We found that M2 population activity during the task could be well modeled by an HMM that established a dictionary between self-initiated actions and neural patterns. To explain the neural mechanism generating reproducible yet temporally variable sequences of patterns, we propose that transitions between attractors are driven by low-dimensional correlated variability. This can be produced by reciprocally connecting a high dimensional recurrent network and a low-dimensional feedforward network. Attractors in the high-dimensional network represent the neural patterns inferred from M2 population activity. Previous experiments showed that recurrent circuits between cortical areas like M2 and subcortical areas such as thalamus \cite{guo2018anterolateral,guo2017maintenance} and basal ganglia nuclei \cite{helie2015learning,desmurget2010motor,nakajima2019prefrontal} are necessary to sustain attractor dynamics and produce motor sequences, and we suggest that cortical-subcortical circuits might correspond to our high- and low-dimensional network interaction. This mechanistic model predicts low-dimensional and sequentially aligned noise correlations, which we confirmed in the empirical data.  While previous work showed that low-dimensional (differential) correlations may be detrimental for accurately encoding external stimuli \cite{moreno2014information}, our results demonstrate that, surprisingly, they are also essential for circuits to produce stable yet temporally variable self-initiated action sequences.

\begin{figure*}[th!]
\captionsetup{justification=raggedright}
\subfloat{\label{fig:1a}}
\subfloat{\label{fig:1b}}
\subfloat{\label{fig:1c}}
\subfloat{\label{fig:1d}}
\subfloat{\label{fig:1e}}
\subfloat{\label{fig:1f}}
\centering
\includegraphics[width=0.9\textwidth]{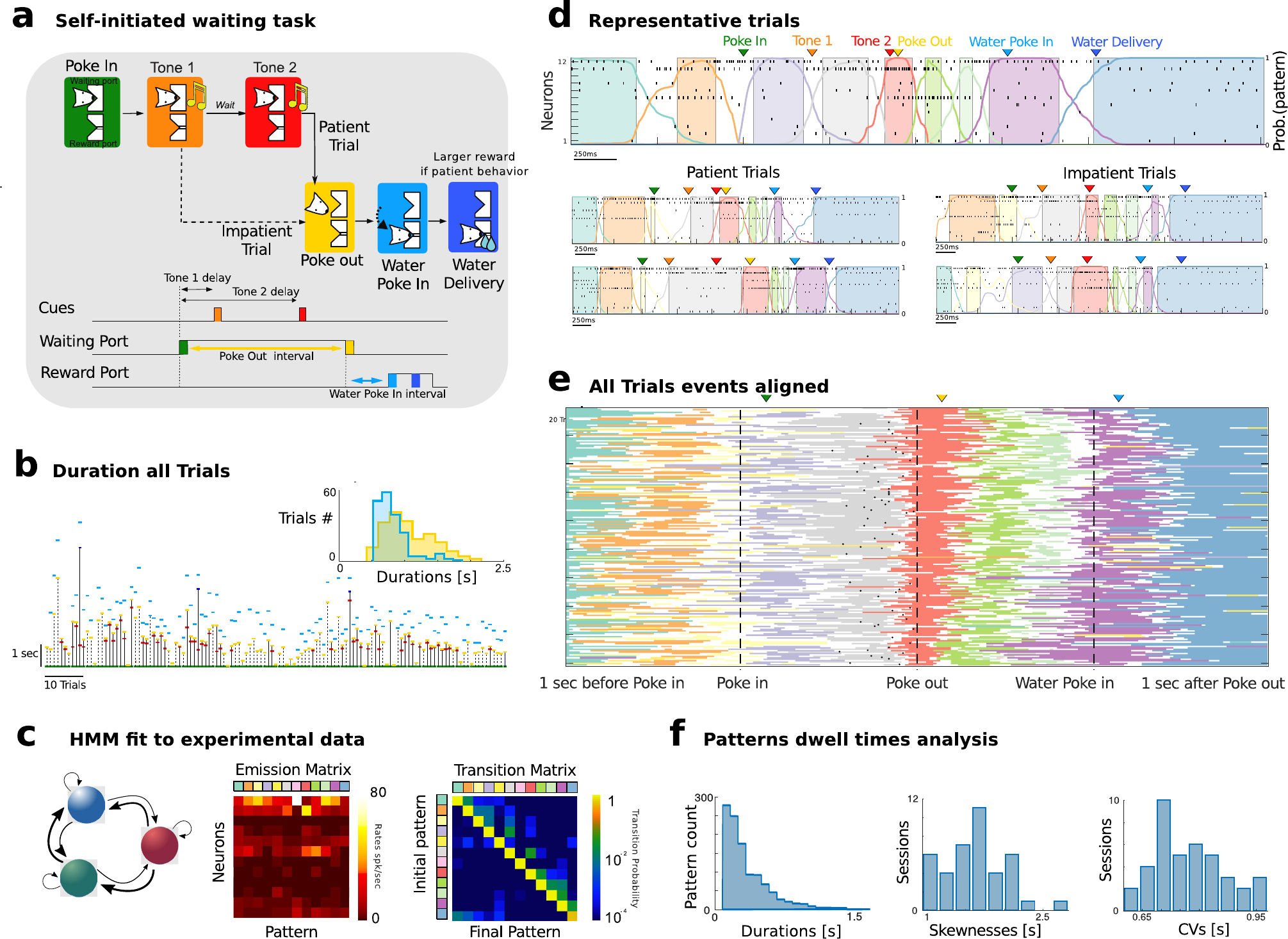}
\centering
\caption{{\bf Waiting task and M2 pattern sequences.} \protect\subref{fig:1a}) Schematic of task events. A rat self-initiated the waiting task by poking into a Wait port (Poke In), where tone 1 was played (after 400 ms), and, after a variable delay, a different tone 2 was played. The animal could decide to Poke Out of the Wait port at any time (after tone 2 in patient trials; between tone 1 and 2 in impatient trials) and move to the Reward port (Water Poke In) to receive a water reward (large and small for patient and impatient trials, respectively). Bottom: schedule of trial events. Three events (PI, PO, WPI) are triggered by self-initiated actions with respective interevents interval highlighted. \protect\subref{fig:1b}) Waiting behavior in a representative session. Vertical bars indicate waiting times for patient or impatient trials (full and dashed lines, respectively). Tick marks represent event times (color-coded as in \protect\subref{fig:1a}). Inset: Interevent interval distribution for self-initiated actions ([PO - PI] and [WPI - PO], yellow and cyan, respectively). \protect\subref{fig:1c}) Neural pattern inference via Hidden Markov Model (HMM). An HMM (left, schematics) is fit to a representative session in \protect\subref{fig:1d}, returning a set of neural patterns (Emission Matrix, center) and a Transition Probability Matrix (TPM, right). Each pattern is a population firing rate vector (columns in the Emission Matrix). The TPM returns the probability for a transition between two patterns to occur. \protect\subref{fig:1d}) Representative trials from one ensemble of 12 simultaneously recorded M2 neurons during patient (top and bottom left) and impatient (bottom right) trials. Top: spike rasters with latent patterns extracted via HMM (colored curves represent pattern posterior probability; colored areas indicate intervals where a pattern was detected with probability exceeding $80\%$). \protect\subref{fig:1e}) All trials from the  representative session (each row corresponds to a trial). Individual trials have been time-stretched to align to five different events (1 s before Poke In, Poke In, Poke Out, Water Poke In, 1 s after Poke Out). All trials display a stereotyped pattern sequence. Color-coded lines represent stretched intervals where patterns were detected (same as colored intervals in \protect\subref{fig:1d}). Black tick marks represent tone 2 onset in patient trials only.
\protect\subref{fig:1f}) Left: Histograms of pattern dwell times across trials in the representative session reveal right-skewed distributions (we excluded the first and last pattern in the sequence, whose duration artificially depends on trial interval segmentation). Skewness and coefficient of variability (CV) of pattern dwell time distributions reveal large trial-to-trial variability (41 sessions). }
\label{fig:1}
\end{figure*}
\FloatBarrier

\section{Results}

\subsection{Ensemble activity in M2 unfolds through reliable pattern sequences}
To elucidate the circuit mechanism underlying self-initiated actions we trained animals on a waiting task. In the waiting task, freely moving rats were trained to perform a sequence of self-initiated actions to obtain a water reward. Animals engaged in the trial by inserting their snout into a wait port, where, after a 400 ms delay, a first auditory tone signaled the beginning of the waiting epoch. Two alternative options were made available: i) waiting for a second tone, delivered at random times, then move to the Reward port to collect a large water amount (henceforth referred to as ``patient" trials); or ii) terminating the trial at any moment before the second tone, then move to a reward port to collect a small amount of water (henceforth referred to as ``impatient" trials). In either case, rewards were collected by withdrawing the snout from the wait port and poking into the reward port; thus, patient and impatient trials shared the same action sequence (\Cref{fig:1a}). The intervals between consecutive actions show large trial-to-trial variability with right-skewed distributions (\Cref{fig:1b} and \Cref{fig:S1a}), suggestive of a potential stochastic mechanism underlying their action timing \cite{killeen1988behavioral}.

To uncover the neural correlates of self-initiated actions, we recorded ensemble spike trains from the secondary motor cortex (M2, from $N=5-20$ neurons per session, $9.1\pm0.5$ on average across 41 recorded sessions) of rats engaged in the waiting task \cite{murakami2014neural,murakami2017distinct}. We found that single-trial ensemble neural activity in M2 consistently unfolded through reliable sequences of hidden or latent neural patterns, inferred via a Poisson hidden Markov model (HMM, see \Cref{fig:1c} and \Cref{fig:S2}). This latent variable model posits that ensemble activity in a given time bin is determined (and emitted) by one of a few unobservable latent activity patterns, represented by a vector of ensemble firing rates (depicted column-wise in the ``emission  matrix''). In the next time bin, the ensemble may either dwell in the current pattern or transition to a different pattern, with probabilities given by rows of the ``transition matrix.'' Stochastic transitions between patterns occur at random times according to an underlying Markov chain, and neurons discharge as Poisson processes with pattern-dependent firing rates. The number of patterns in each session was selected via an unsupervised cross-validation procedure ($10.3\pm 4.1$ across 41 sessions, which ranged from 4 to 21 patterns, \Cref{fig:S2}; see Methods). The identity and order of inferred patterns were remarkably consistent within each session even across patient and impatient trials, (\Cref{fig:1e} and \Cref{fig:S3}). The average pattern dwell time was $0.50\pm0.28$ s (\Cref{fig:1f}), in agreement with previous findings in other cortical areas \cite{mazzucato2015dynamics}. Such long dwell times, which are greater than typical single neuron time constants, suggest that the observed patterns may be an emergent property of the collective circuit dynamics within M2 and reciprocally connected brain regions. Crucially, even though the identity and order of patterns within a sequence were highly consistent across trials, pattern dwell times showed large-trial to trial variability, characterized by right-skewed distributions (\Cref{fig:1f}, coefficient of variability CV=$0.78\pm 0.10$ and skewness $1.67\pm 0.46$). This temporal heterogeneity suggests that a stochastic mechanism may contribute to driving transitions between consecutive patterns within a sequence.

\FloatBarrier
\begin{figure*}[th!]
\captionsetup{justification=raggedright}
\subfloat{\label{fig:2a}}
\subfloat{\label{fig:2b}}
\subfloat{\label{fig:2c}}
\subfloat{\label{fig:2d}}
\centering
\includegraphics[width=1.\textwidth]{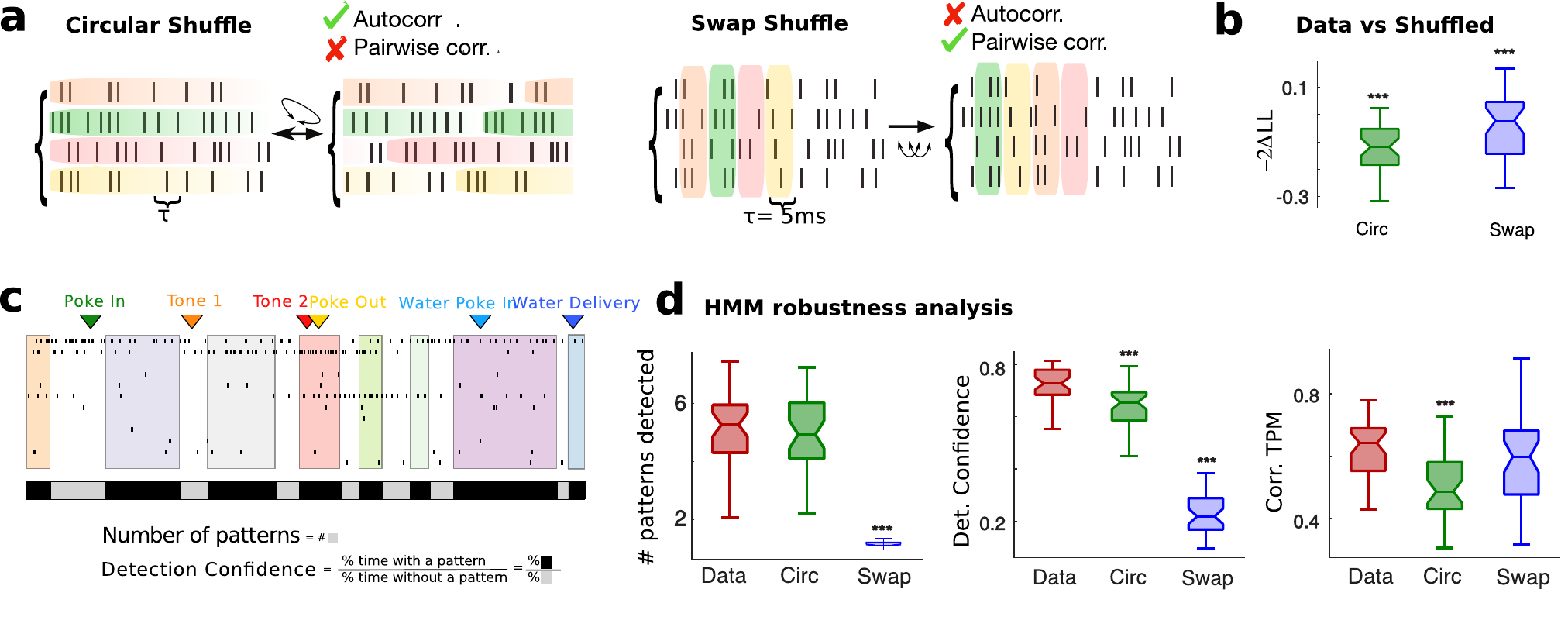}
\centering
\caption{{\bf Robustness of pattern inference.} \protect\subref{fig:1a}) Schematic of shuffled procedure to create surrogate datasets: Circular Shuffle (left) preserved single-cell autocorrelations and destroyed pairwise correlations; Swap Shuffle (right) preserved pairwise correlations and destroyed autocorrelations. \protect\subref{fig:1b}) Difference in $-2\times$Log-likelihood between HMM fit of empirical dataset and of the two surrogate datasets. \protect\subref{fig:1c}) Representative trial showing detection confidence measure (same color-coded notation as in \Cref{fig:1}; black and grey bars: fraction of trial duration where patterns were detected with probability larger or smaller than $80\%$, respectively.  \protect\subref{fig:1d}) HMM robustness analyses. Left: Number of patterns detected across sessions for empirical and surrogate datasets. Center: Pattern detection confidence, estimated as fraction of trials were patterns were detected with probability exceeding $80\%$. Right: Consistence of pattern sequence, estimated as Pearson correlation coefficients between single-trial estimates of ``symbolic'' TPMs encoding the sequence identity (see Methods section \ref{methods:comparison}). \protect\subref{fig:1b}-\protect\subref{fig:1d}: signed-rank tests between empirical and shuffled datasets, $*=p<0.05$, $**=p<0.01$, $***=p<0.001$. }
\label{fig:2}
\end{figure*}

\subsection{Pattern sequences capture dynamics beyond auto- and pairwise correlations}

Before investigate this mechanism (section 2.5, below), we show a number of analyses directed at quantifying how well the HMM-inferred pattern sequences captured ensemble spiking activity beyond single-cell autocorrelations and pair-wise correlations. To do so, first we performed a cross-validation analysis comparing the data to two surrogate datasets (\Cref{fig:2a}) \cite{maboudi2018uncovering}. In the ``circular-shuffled'' surrogate dataset, we circularly shifted bins for each neuron within a trial (i.e., row-wise), thus destroying pairwise correlations but preserving single-cell autocorrelations. In the ``swap-shuffled'' surrogate dataset, we randomly permuted  population activity across bins within a trial (i.e. column--wise), thus preserving instantaneous pairwise correlations but destroying autocorrelations. We found that the cross-validated likelihood of held-out trials for an HMM trained on the real dataset was significantly larger compared to an HMM trained on surrogate datasets (\Cref{fig:2b}, empirical vs. circular shuffled: $p=9\times 10^{-8}$; vs. swap shuffled:  $p=0.051$, signed-rank test). When we destroyed autocorrelations, the model entirely failed to detect pattern transitions, leaving only one pattern (\Crefrange{fig:2c}{fig:2d}, $p=2.4\times 10^{-8}$). When destroying pairwise correlations, the model still detected multiple patterns whose number was in the same range as the model trained on the empirical data (\Cref{fig:2d}, $p=0.11$). However, pattern detection was significantly less confident than in empirical data (\Cref{fig:2c,fig:2d}, $p=1.4\times 10^{-7}$); moreover, inferred pattern sequences were significantly sparser and more similar across trials in the data compared to the surrogate datasets (\Cref{fig:2d}, $p=9\times 10^{-8}$, \Cref{fig:S2c}-\Cref{fig:S2d}). We concluded that pattern sequences captured the single-trial dynamics of population activity beyond autocorrelations and pairwise correlations.

\subsection{Patterns arise from dense and distributed neural representations}

\begin{figure*}[th!]
\captionsetup{justification=raggedright}
\subfloat{\label{fig:3a}}
\subfloat{\label{fig:3b}}
\subfloat{\label{fig:3c}}
\subfloat{\label{fig:3d}}
\subfloat{\label{fig:3e}}
\subfloat{\label{fig:3f}}
\centering
\includegraphics[width=1.\textwidth]{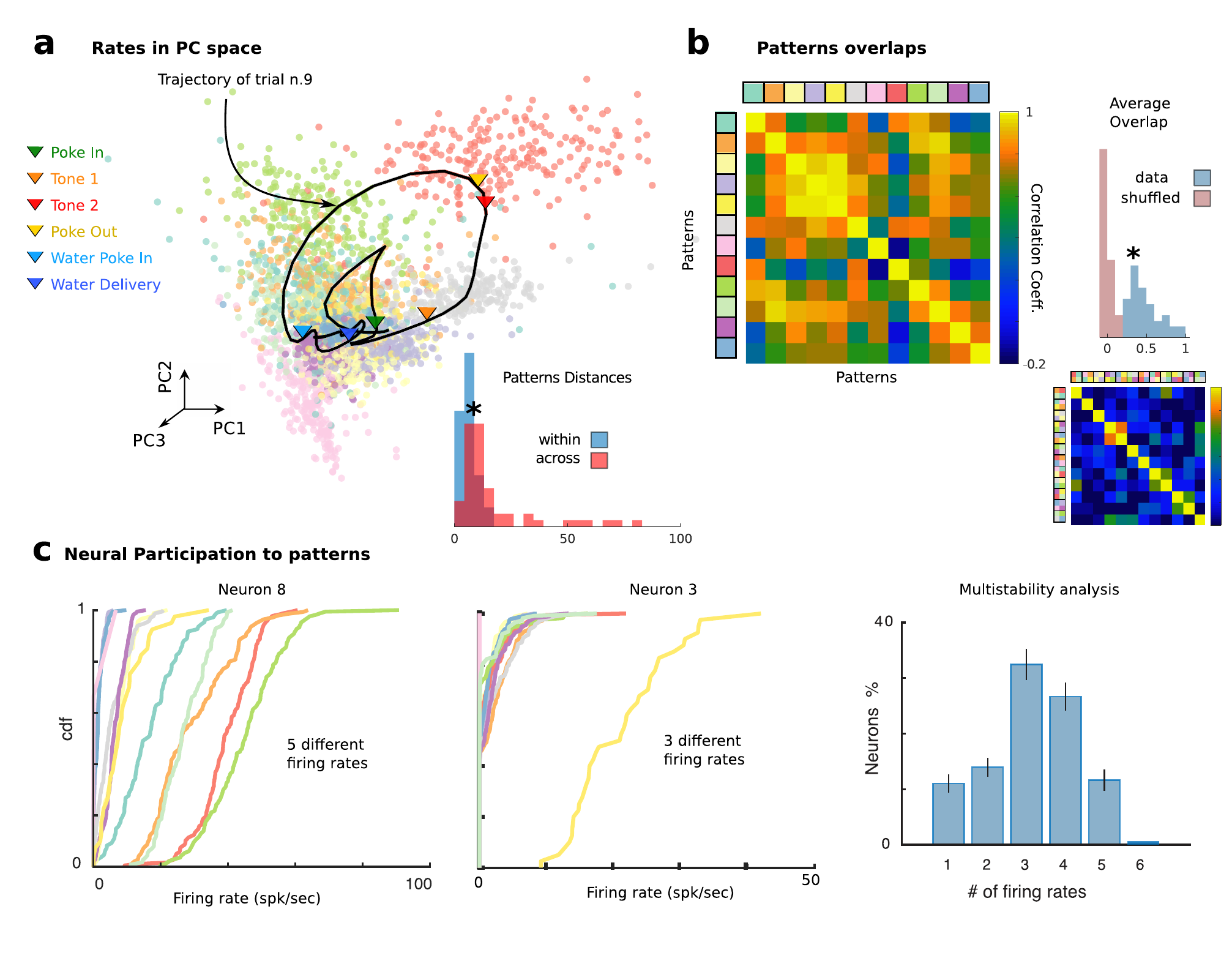}
\centering
\caption{{\bf Dense and distributed population code in M2.} \protect\subref{fig:3a}) Neural patterns cluster in Principal Component space (all trials from the representative session in \Cref{fig:1}; color-coded dots represent patterns in single trials; one representative trial smoothed trajectory overlaid where arrows show events onsets along trajectory). Inset: Distribution of within- and across-cluster distances between patterns (ranksum test $p<2.0\times 10^{-7}$) \protect\subref{fig:3b}) Pearson correlation matrix between patterns reveals significantly larger overlaps in the empirical data (top left: representative session) compared to those found when drawing random patterns from the empirical firing rate distribution (bottom right). Inset: Distribution of pattern correlations for empirical (blue) versus shuffled datasets (red). \protect\subref{fig:3c}) Single neuron firing rates are modulated by pattern sequences.
Left: Cumulative firing rate distributions conditioned on patterns (color-coded as \protect\subref{fig:3a}) and \Cref{fig:1d}) for two neurons from the representative ensemble, revealing 5 and 2 significantly different firing rates across patterns, respectively (see Methods section \ref{methods:multistability}. Right: Number of different firing rates per neuron revealed multistable dynamics where $88\pm 2\%$ of neurons had activities modulated by patterns.}
\label{fig:3}
\end{figure*}

How do pattern sequences emerge from neural activity? Patterns formed separate clusters tiling population activity space, with between-cluster distances being significantly larger than within-cluster distances (\Cref{fig:3a}, Wilcoxon rank-sum test, $p<10^{-20}$). Most neurons were active in several patterns, leading to dense neural representations, where overlaps between patterns (0.41$\pm$0.22, defined as Pearson correlation between firing rate vectors) were significantly larger than expected solely on the basis of the underlying firing rate distribution (\Cref{fig:3b}, $p<3\times 10^{-18}$, t-test). We found that the vast majority of neurons (88$\pm$2$\%$) were  ``multistable'', with their firing rates significantly modulated across patterns (\Cref{fig:3c}), in agreement with previous findings \cite{mazzucato2015dynamics}. In particular, neurons attained on average 3.2$\pm$1.2 different firing rates across patterns. We concluded that most M2 neurons participated in the pattern sequences, suggesting that M2 neural populations can support dense and distributed representations characterized by mixed selectivity to multiple patterns. 

\FloatBarrier

\subsection{Pattern onsets predict self-initiated actions}

What kind of information about self-initiated behavior can be decoded from M2 pattern sequences? The statistical structures of neural patterns and action sequences shared remarkable similarities: single-trial consistency of identity and order of actions/patterns within a session, yet right-skewed distributions of timing intervals across trials (\Cref{fig:1} and \Cref{fig:S3}). We thus hypothesized that the onset of specific neural patterns could be causally involved with and therefore predictive of the timing and identity of upcoming self-initiated actions. 

To test this hypothesis, we aimed to establish a cross-validated dictionary between actions and neural patterns, which we did by tagging the onset of specific patterns with the actions they most strongly predicted (\Cref{fig:4a}). This automated tagging method showed that, even though both pattern onsets and actions occurred at highly variable times in different trials, action onset times were reliably preceded by specific patterns onset on a sub-second scale ($168\pm175$ ms, interval between pattern onset and tagged action). To assess the significance of the action/pattern dictionary, we tested whether pattern onsets could correctly predict actions performed during incorrect trials ($32.5\pm15.8\%$, see \Cref{fig:4b} and \Cref{fig:S4}), based on the dictionary learned solely from correct trials (defined as those in which a visit to the waiting port was following directly by a movement to the reward port; both patient and impatient types, $67.4\pm15.8\%$ fraction of trials per session, see \Cref{fig:1a}); other action sequences, where the animal behaved erratically, were deemed as incorrect trials (\Cref{fig:S4}). When using the cross-validated action/pattern dictionary learned on correct trials, we were able to correctly predict which actions the animal would perform in incorrect trials (\Crefrange{fig:4b}{fig:4c}). 

We then restricted our attention to correct trials and further reasoned that, because correct trials entailed the same set of actions in both patient and impatient conditions, if patterns encoded upcoming actions (as oppose to, e.g., reward size expectations) then pattern occurrence would be indistinguishable between the two kinds of trials. We indeed confirmed that the distribution of pattern dwell times was not significantly different between these two conditions in $95\%$ of sessions (Kolmogorov-Smirnov test, $p>0.05$). We thus concluded that the spatiotemporal variability observed in M2 population activity in single trials is consistent with a mechanism whereby specific pattern onsets anticipate self-initiated actions because they are causally upstream, as expected of motor regions (although we did not establish causality).

\subsection{Correlated variability generates sequences of metastable attractors}

What is a possible circuit mechanism underlying the observed pattern sequences? We aimed at capturing three main features of the empirical behavioral and neuronal data: ({\small \textsc{i}}) Patterns were long lived (0.5 s on average, \Cref{fig:1f}), more than one order of magnitude longer than single neuron time constants, suggesting they originate in attractor dynamics as an emergent network property. ({\small \textsc{ii}}) Pattern dwell time distributions were variable and right-skewed (see \Cref{fig:1f}), suggesting that transitions between patterns may be noise-driven (see e.g. \cite{gardiner1985handbook}). ({\small \textsc{iii}}) Sequences of patterns were highly reliable, the same sequence consistently recurring across trials (\Cref{fig:1e} and \Cref{fig:S3}). 
We thus wished to construct a mechanistic model generating reliable sequences of long-lived attractors, where transitions between attractors were driven by noise.  First, we demonstrate a computational mechanism explaining the empirical features of the observed sequences, then we map the model components onto a mesoscale circuit, and finally we will test model predictions with experimental data. 

The crucial ingredient driving transitions between patterns in the model entails constraining population activity fluctuations along a low-dimensional manifold within a high-dimensional of activity space. We achieved this by embedding a low-rank term in the synaptic couplings, whose structure was as follows. 

We modeled population activity in M2 as arising from a recurrent network of rate units governed by the following dynamics:
\begin{equation}
\label{model:1results}
 \tau \dot{u}_i(t)=-u_i(t)+\sum_{j=1}^{N}J^S_{ij}\phi_j(u_j(t))+\zeta(t) \sum_{j=1}^NJ^F_{ij}\phi(u_j(t)),
\end{equation}
where $u_i$ and $\phi_i(u_i)$ are post-synaptic currents and single-neuron current-to-rate transfer functions representing the activity of M2 neurons. Current-to-rate functions in the model were fit to those estimated from the empirical data in M2 (see \Cref{fig:5a}, \Cref{fig:S5}, Methods and \cite{lim2015inferring}).
We hypothesized that patterns originated from $p$ discrete attractors $\eta^{\mu}$, for $\mu=1,\ldots,p$, stored in the recurrent synaptic couplings $J^S_{ij}\propto \sum_{\mu=1}^{p} f[\eta^{\mu}_i]g[\eta^{\mu}_j]$, connecting a pre-synaptic neuron $i$ and a post-synaptic neuron $j$ in M2 ($f$ and $g$ are threshold functions, see Methods and \cite{pereira2018attractor}). This is consistent with experimental evidence supporting discrete attractor dynamics in secondary motor cortex \cite{schmitt2017thalamic,guo2017maintenance,inagaki2019discrete}. Because we sought to generate transitions stochastically, the model operates in a regime where the attractors $\eta^{\mu}$ were stable in the absence of the second term $J^F$  (\Cref{fig:S6a}). Transitions between attractors, giving rise to sequences, originate from the asymmetric  term $J^F_{ij}\propto \sum_{\mu=1}^{p} f[\eta^{\mu+1}_i]g[\eta^{\mu}_j]$ in Eq.~(\ref{model:1results}), henceforth referred to as {\it correlated variability} term. This term generates stochastic dynamics via the noise $\zeta(t)$, with mean $\bar \zeta$ and variance $\sigma_\zeta^2$. We will discuss below the mechanistic origin of this term.

The correlated variability term constrains population activity fluctuations onto a low-dimensional manifold within activity space, whose dimension is bounded by the number $p$ of attractors, thus much smaller than the number of neurons $N$. The effect of this term is to generate population activity fluctuations which are correlated across neurons. We found that, within a large range of parameters (\Cref{fig:S6b}), the network model met all our objectives: ({\small \textsc{i}}) the model generated long-lived attractors (0.98$\pm$1.19s, \Cref{fig:5b}), whose duration was orders of magnitude longer than single-neuron time constants ($\tau=20$ms, see Table~\ref{table:1}), thus emerging from the network's collective dynamics. ({\small \textsc{ii}}) Crucially, dwell time distributions were right-skewed with large coefficient of variability (\Cref{fig:5c}), capturing the large trial-to-trial variability observed in the empirical distributions of behavioral and neural data (\Cref{fig:1}). Since attractors would be stable in the absence of noise $\zeta(t)$ ((\Cref{fig:S6a}), transitions between attractors were entirely noise-driven in this model. ({\small \textsc{iii}}) Despite the variability of timing, the sequence of attractors was highly reliable, as in the empirical behavioral and neural data.

Furthermore, we found that single-neuron firing rate distributions were heterogeneous (\Cref{fig:5d}), similar to the empirical ones (\Cref{fig:3c}). In particular, most neurons participated in the sequential dynamics, attaining on average $3.8\pm0.9$ different firing rates across patterns, explaining the single-neuron multistability properties observed in M2 neural data (\Cref{fig:5c}, see also \cite{mazzucato2015dynamics}). We conclude that metastable attractor dynamics in our model captured the lexically stable yet temporally variable features of pattern sequences observed in the empirical data.

\subsection{Correlated variability originates in a mesoscale feedback loop}

The crucial ingredient driving transitions between patterns in the model (see Eq.~(\ref{model:1results})) entails restricting fluctuations along a low-dimensional manifold within activity space. We achieved this by embedding a low-rank noise term in the synaptic connectivity architecture of the neural circuit. What is the circuit origin of these couplings? We found that this low-rank structure naturally arises from a two-area model, describing a feedback loop between a large recurrent circuit representing M2 and a small feedforward circuit (provisionally denoted as Y):
\begin{eqnarray}
\label{model:results1}
 \tau \dot{u}_i(t)&=&-u_i(t)+\sum_{j=1}^{N}J^S_{ij}\phi_j(u_j(t))+\sum_{j=1}^{N_Y}W^{(M2 \leftarrow Y)}_{ij}r_j
 \ ,\\
 \tau_{Y} \dot{r}_i&=& - r_i + \sum_{j=1}^NW^{(Y\leftarrow M2)}_{ij}\phi_j(u_j)\ .\nonumber
\end{eqnarray}
Here, $u_i$ represent the activity of M2 neurons (same as in Eq. \ref{model:1results}), and $r_i$ represent activities of neurons in area Y. The latter area is smaller ($N_Y\ll N$) and faster ($\tau_Y<\tau$), and lacks recurrent couplings, suggesting it may correspond to a subcortical circuit. The asymmetric term $J^F$ in Eq.~(\ref{model:1results}), which generates stochastic transitions between otherwise stable M2 attractors, originates from the reciprocal couplings $W^{(Y \leftarrow M2)}$, $W^{({M2}\leftarrow Y)}$ between M2 and area Y in Eq.~(\ref{model:results1}), its temporal dependence arising from short-term plasticity at these synapses (see Methods \cite{tsodyks1998neural}). The reciprocal connections $W$ in this two-area model can be integrated out when dynamics in area Y are faster than in M2 ($\tau_{Y}<\tau$) \cite{reinhold2015distinct,jaramillo2019engagement}. The two-area mesoscale model in Eq.~(\ref{model:results1}) is then mathematically equivalent to the effective dynamics in Eq.~(\ref{model:1results}), whose recurrent couplings are augmented to include an asymmetric term $J^F$, inherited from the reciprocal loop. In the Methods section we show how the mean and variance of the noise term $\zeta(t)$ in Eq.~(\ref{model:1results}) capture, respectively, the strength and the variability of the couplings in the feedback loop between M2 and area Y. Its time dependence arises from short term plasticity in these couplings assuming area Y is small. 

\subsection{\label{section:private}Correlated variability is necessary to explain temporal variability}

Is it possible to generate the observed pattern sequences with alternative mechanisms, in the absence of correlated variability? We tested a large class of models where synaptic couplings included both symmetric ($J^{S}$) and asymmetric ($J^{F}$) terms, but where the asymmetric couplings were constant, i.e., setting $\sigma_\zeta=0$ in Eq.~(\ref{model:1results}). Depending on different ratios between asymmetric to symmetric couplings, this class of models generated either decaying activity, or stable attractors, or sequences of attractors when the ratio was large enough (\Cref{fig:S6a}). However, in the whole region of parameter space with sequential dynamics, this class of models failed to capture crucial aspects of the data. Namely, dwell times distributions were short and they showed no trial-to-trial variability, thus being incompatible with the observed patterns (\Cref{fig:1f}). 

We then attempted to rescue these models by driving the network with increasing levels of {\it private noise}, namely, external noise, independent for each
neuron (\Cref{fig:S7a}, see Methods). This led to small amounts of trial-to-trial variability in dwell times, but was still qualitatively different from the empirical data. Increasing the private noise level beyond a critical value destroyed sequential activity (\Cref{fig:S7b}).

We reasoned that the difficulty in generating long-lived, right-skewed distributions of dwell times in this alternative class of models was due to the fact that transitions were not driven by noise, but by the deterministic asymmetric term $J^{F}$.
Adding private noise did not qualitatively change variability, due to the high dimensionality of the stochastic component. Private noise induces independent fluctuations in each neuron; however, in order to drive transitions from one attractor to the next one within a sequence, these fluctuations have to align along one specific direction in the N-dimensional space of activities. The probability that independent fluctuations align in a specific direction vanishes in the limit of large networks, explaining why in the private noise model transitions cannot driven by noise. We thus concluded that correlated variability was necessary to reproduce the right-skewed distribution of pattern dwell time observed in the data. 

\FloatBarrier
\begin{figure*}[th!]
\captionsetup{justification=raggedright}
\subfloat{\label{fig:4a}}
\subfloat{\label{fig:4b}}
\subfloat{\label{fig:4c}}
\subfloat{\label{fig:4d}}
\subfloat{\label{fig:4e}}
\subfloat{\label{fig:4f}}
\centering
\includegraphics[width=1.\textwidth]{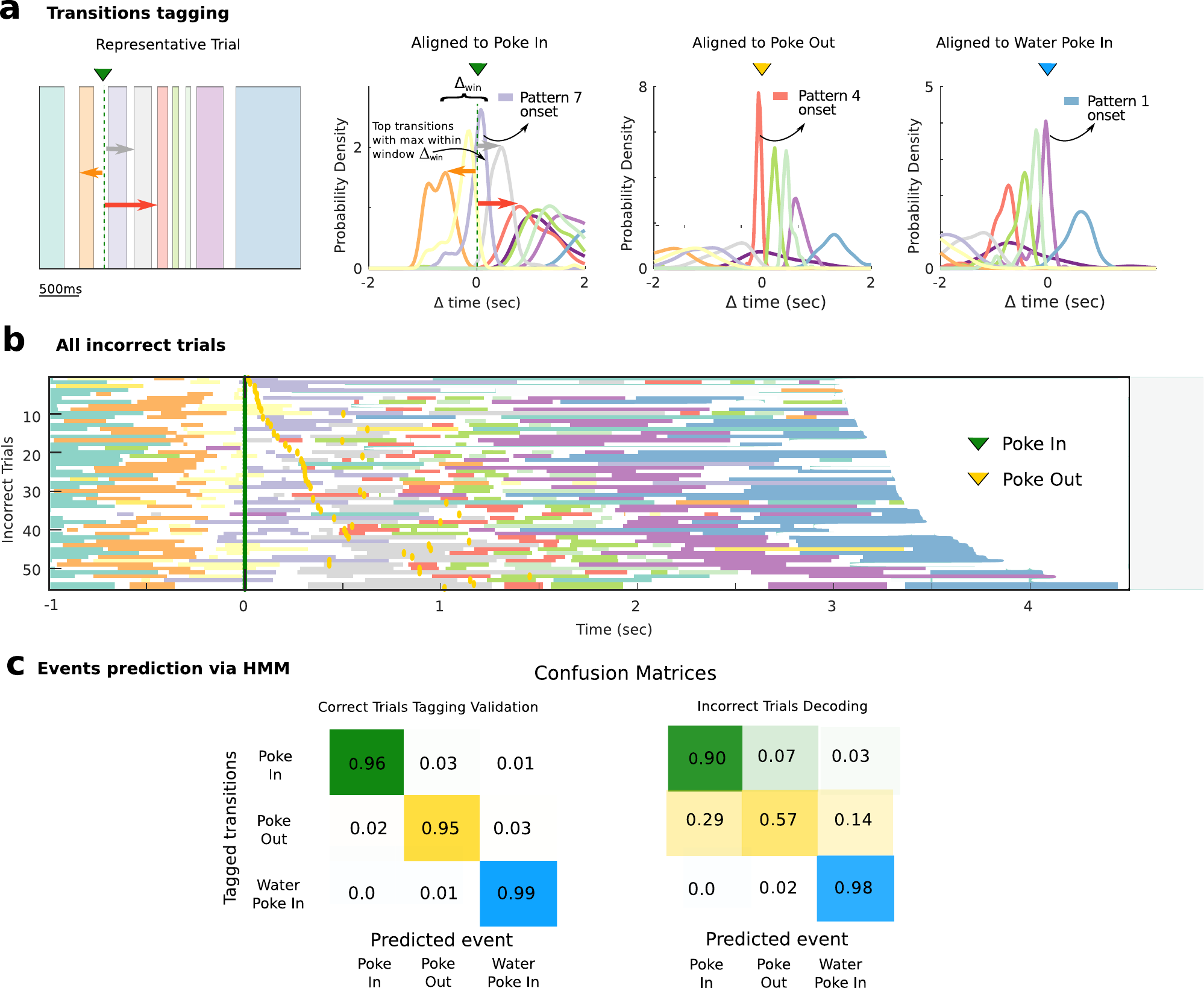}
\centering
\caption{{\bf Predicting self-initiated actions from neural pattern onsets} \protect\subref{fig:4a}) Schematic of pattern/action dictionary. Left: For each action in a correct trial (left: representative trial from \Cref{fig:1d}), pattern onsets are aligned to that action (Poke In in this example). The pattern whose median onset occurs within an interval $\Delta_{win}=[-0.5,0.1]$ s aligned to the action, and whose distribution has the smallest dispersion, is tagged to that action (color-coded curves are distributions of action-aligned pattern onsets from all correct trials in the representative session in \Cref{fig:1}). \protect\subref{fig:4b}) In incorrect trials (55 trials from the same representative session; time $t=0$ aligned to Poke In), the same patterns as in correct trials are detected (cfr. \Cref{fig:1e}), but they concatenate in different sequences. \protect\subref{fig:4c}) Predicting self-initiated actions from pattern onsets. Left panel: In correct trials (split into training and test sets), using a pattern-action dictionary established on the training set (procedure in panel a), action onsets are predicted on test trials (confusion matrix: cross validated prediction accuracy averaged across 41 sessions; hits: correct action predicted within an interval of $[-0.1,0.5]$s aligned to pattern onset). Right panel: In incorrect trials, actions onsets are predicted based on the cross-validated dictionary established in correct trials.}
\label{fig:4}
\end{figure*}

\FloatBarrier
\begin{figure*}[th!]
\captionsetup{justification=raggedright}
\subfloat{\label{fig:5a}}
\subfloat{\label{fig:5b}}
\subfloat{\label{fig:5c}}
\subfloat{\label{fig:5d}}
\subfloat{\label{fig:5e}}
\subfloat{\label{fig:5f}}
\subfloat{\label{fig:5g}}
\centering
\includegraphics[width=1.\textwidth]{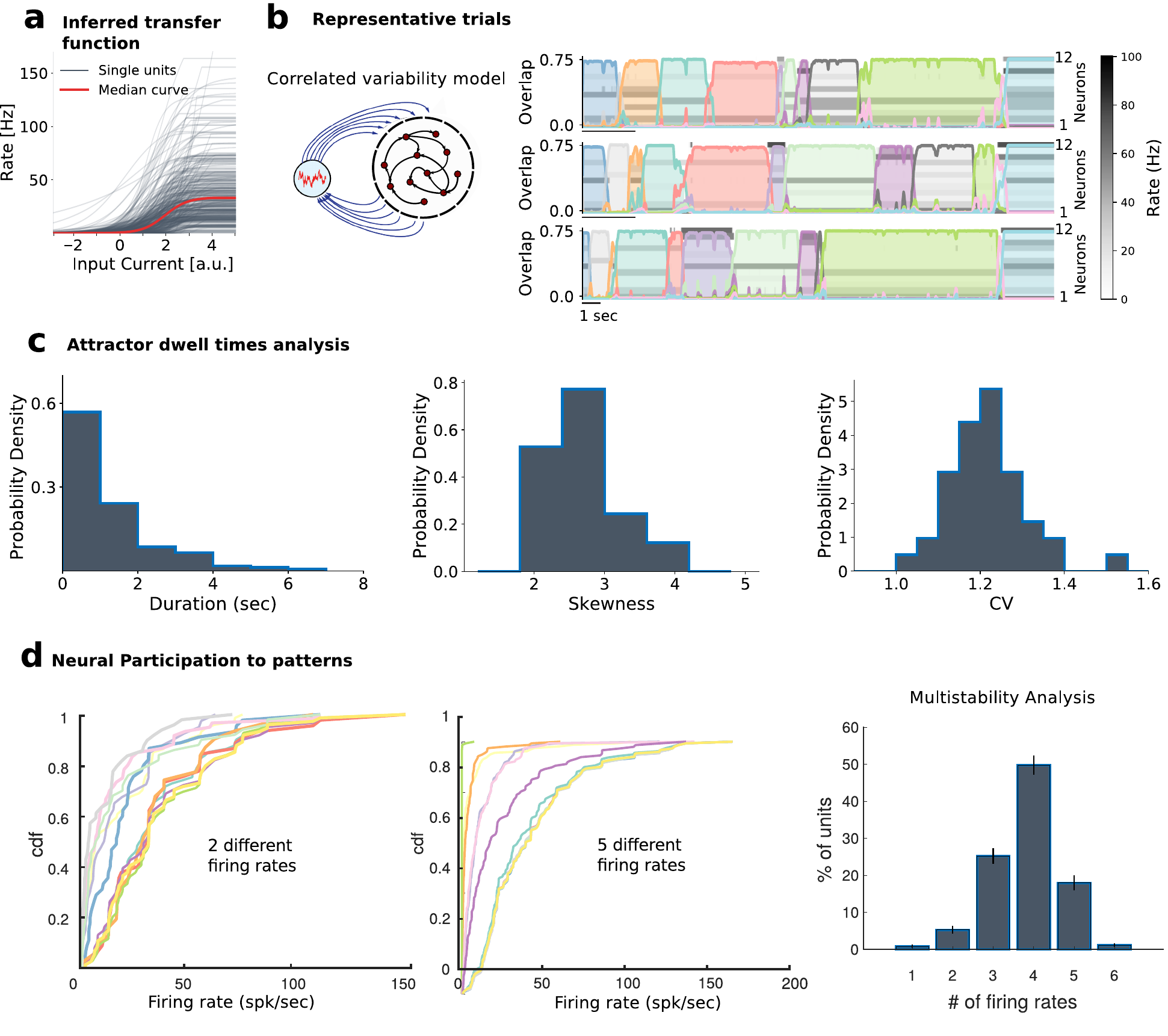}
\centering
\caption{{\bf Attractor model of pattern sequences.} \protect\subref{fig:5a}) Distribution of empirical single-cell current-to-rate transfer functions $\phi_i$ inferred from the data (366 neurons from 41 sessions), used as transfer functions in the recurrent network model (see Methods). \protect\subref{fig:5b}) The correlated variability model (Eq.~\ref{model:results1}) generates reliable sequences of long-lived attractors with large trial-to-trial variability in attractor dwell times (representative trials: rows represent the activity of 12 neurons randomly sampled from the network; color-coded curves represent time course of overlaps (see Eq.~\ref{model:overlap}) between population activity and each attractor; detected attractors are color-shaded). \protect\subref{fig:5c}) Histogram of attractor dwell times across trials in the representative network of \protect\subref{fig:5b}) reveals right-skewed distributions (left, we excluded the first and last pattern in the sequence, whose duration artificially depends on trial interval segmentation). Skewness (center) and coefficient of variability (CV, right) of pattern dwell time distributions reveal large trial-to-trial variability (41 simulated networks).
\protect\subref{fig:5d}) Single neuron firing rates are modulated by pattern sequences in the model.
Cumulative firing rate distributions conditioned on attractors (color-coded as in \protect\subref{fig:5b})) for two representative neurons in the model, revealing 2 and 3 significantly different firing rates across attractors, respectively (see Methods section \ref{methods:multistability}). Inset: Number of different firing rates per neuron revealed multistable dynamics where $99\pm 1\%$ of neurons had activities modulated by patterns.}
\label{fig:5}
\end{figure*}

\subsection{Low-dimensional variability of M2 attractors dynamics}

\begin{figure*}[th!]
\captionsetup{justification=raggedright}
\subfloat{\label{fig:6a}}
\subfloat{\label{fig:6b}}
\subfloat{\label{fig:6c}}
\centering
\includegraphics[width=1.\textwidth]{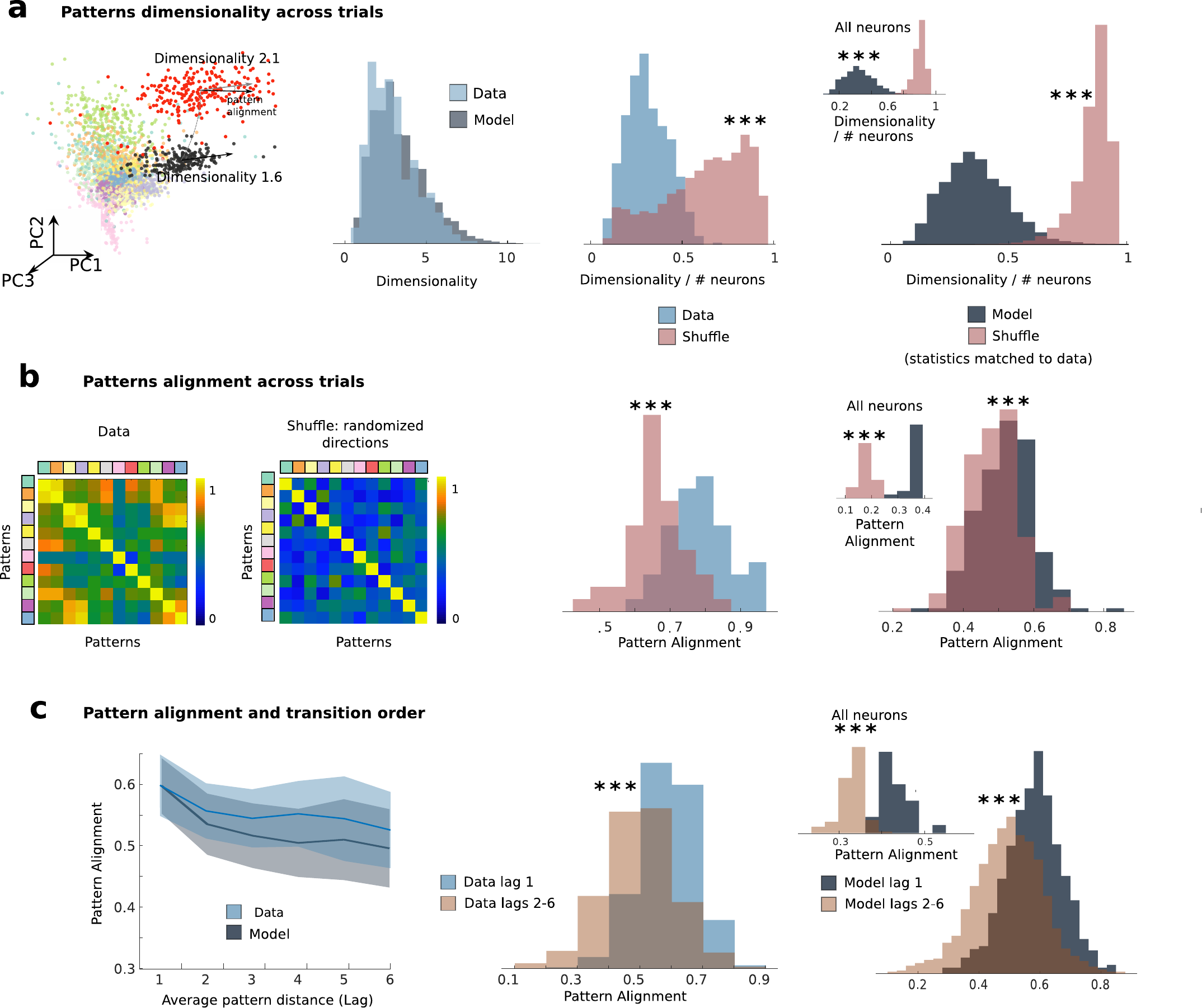}
\centering
\caption{{\bf Low-dimensional variability in models and data.}  \protect\subref{fig:6a}) Comparison of dimensionality of pattern-conditioned noise correlations in the data (blue) and the model (grey) reveals low-dimensional population activity fluctuations, significantly smaller than expected by chance (red, shuffled datasets). From left to right: first panel, representative session as in \Cref{fig:1}); second panel, summary across 41 sessions from the data and the model; third panel, fractional dimensionality in the data; fourth panel: model dimensionality estimated by matching ensemble sizes and number trials to data across 41 simulated sessions; inset: dimensionality estimated from N=10000 neurons in 41 simulated sessions. \protect\subref{fig:6b}) Pattern-conditioned noise correlations are highly aligned between patterns in the data. Alignment between top canonical correlation vectors (blue, data; gray, model) is larger than between random principal component directions (red). \protect\subref{fig:6c}) Left: Alignment of noise correlations between each pattern and patterns occurring at lag $n$ in the sequence (e.g., $n=1$ represents patterns immediately preceding or following the reference pattern) in the model (grey) and in the data (blue). Pattern alignments are significantly larger for patterns at one lag compared to patterns at longer lags.  
All panels: $*=p<0.05$, $***=p<0.001$.}
\label{fig:6}
\end{figure*}

Our recurrent network model (see Eq.~(\ref{model:1results})) entails a specific hypothesis for the mechanism underlying the observed sequences: transitions between consecutive attractors are generated by correlated variability. We reasoned that, if this was the mechanism at play in driving sequences, then two clear predictions should be borne out in the neural population data. First, the correlated variability term in Eq.~(\ref{model:1results}) predicts that population activity fluctuations within a given attractor (color-shaded intervals in \Cref{fig:5b}), henceforth referred to as ``noise correlations'', lie within a subspace whose dimension is much smaller than that expected by chance (\Cref{fig:6a}, dimensionality in the model vs. shuffled surrogate dataset, ranksum test, $p<10^{-15}$). Second, the sequential structure of the correlated variability term in Eq.~(\ref{model:1results}) implies that noise correlation directions for attractors that occur in consecutive order within a sequence should be co-aligned. A canonical correlation analysis showed that in the correlated variability model the alignment across attractor (measured using the top $K$ principal components of the noise correlations, where $K$ is its dimensionality) was much larger than expected by chance (\Cref{fig:6b}, alignment in the model vs. shuffled surrogate dataset, ranksum test, $p<10^{-5}$). More specifically, we found that the strongest alignment occurred between consecutive attractors within a sequence, compared to attractors occurring further apart (\Cref{fig:6b}, ranksum test, $p<10^{-20}$). 

Having established strong statistical features regarding low-dimensional, aligned noise correlations structure, we tested whether the structure predicted by the model were observed in the M2 neural ensemble data. We defined noise correlations in the empirical data as population activity fluctuations around each neural pattern inferred from the HMM fit (\Cref{fig:6a}). Applying the same analyses to the data that were run on the model, we found that indeed empirical noise correlations around each neural pattern had lower dimension than expected by chance, and closely matched the dimensionality predicted by the model (\Cref{fig:6a}). CCA further revealed that noise correlations were highly aligned between patterns, significantly above the alignment expected by chance (\Cref{fig:6b}, rank-sum test $p=1.70\times 10^{-4}$). Finally, directions of variability were more aligned between consecutive patterns, compared to patterns further apart in the sequence, (rank-sum test, $p<10^{-14}$; see \Cref{fig:6c}). Thus, the features of the noise correlations in the neural ensemble data were strongly consistent with the predictions from the correlated variability model. 

\FloatBarrier

\section{Discussion}

Our results establish a correspondence between self-initiated actions and attractor dynamics in secondary motor cortex (M2). We found that population activity in M2 during a self-initiated waiting task unfolded through a sequence of patterns, with each pattern reliably predicting the onset of upcoming actions. We interpreted the observed patterns as metastable attractors emerging from the recurrent dynamics of a two-area neural circuit. The model was capable of robustly generating reliable sequences of metastable attractors recapitulating the properties of the dynamics found in the empirical behavioral and neural data. We propose a neural mechanism explaining the variability in attractor dwell times as originating from correlated variability in a two-area model. The model predicts that population activity fluctuations around each attractor (i.e., ``noise correlations") are highly aligned between attractors and constrained to lie on a low-dimensional subspace, and we confirmed these predictions in the empirical neural (M2) data. Our work establishes a mechanistic framework for investigating the neural underpinnings of self-initiated actions and demonstrates a novel link between correlated variability and attractor dynamics.

\subsection{Evidence for discrete attractor dynamics in cortex}

Evidence from primate and rodent studies supports the hypothesis that cortical circuits generate discrete activity patterns, interpreted as attractors, during working-memory and decision-making \cite{cueva2018delay}. Attractors are characterized by long periods where neural ensembles discharge persistently at approximately constant firing rate (defining a neural pattern) punctuated by relatively abrupt transitions to a different relatively constant pattern. Selective persistent activity during delay periods has been reported in temporal \cite{fuster1981inferotemporal,miyashita1988neuronal} and frontal areas  \cite{fuster1971neuron, funahashi1989mnemonic} in primates, and frontal areas in rodents \cite{erlich2011cortical,schmitt2017thalamic,guo2017maintenance,inagaki2019discrete}. Evidence for attractors encoding sensory stimuli was found in rodent sensory cortex \cite{Jones2007,PonceAlvarez2012,mazzucato2015dynamics}. Optogenetic stimulation of few neurons within an ensemble was shown to drive sustained activation of the entire ensemble \cite{marshel2019cortical}, persisting for seconds after stimulation offsets  \cite{wei2019orderly}, compatible with predictions from attractor models \cite{Amit1997a}. 

Attractor dynamics may depend on the precise temporal structure of the task: in similar tasks, depending on whether the delay period was of fixed or randomized length, either discrete attractors or ramping activity were observed \cite{inagaki2019discrete}. Our waiting task differs from these delayed memory tasks in two respects. First, we analyzed the entire p-step action sequence of freely moving animals, rather than a single step of behavior (i.e. delay period). Second, many steps within the action sequence were self-initiated, rather than prompted by experimenter-controlled signals. Consequently, we uncovered a new dynamical regime in which (i) corresponding behavioral and neural states were metastable, with large trial-to-trial variability in dwell times (ranging from hundreds of ms to a few seconds); ii) states were concatenated into a sequence, reliably occurring in most trials, in accordance with the actual action sequence. Experimental evidence for stimulus-specific sequences of metastable attractors was previously found in primate frontal areas \cite{gat1993statistical,Abeles1995a,Seidemann1996} and rodent sensory areas \cite{Jones2007}. Random sequences were also observed during ongoing periods \cite{mazzucato2015dynamics,mazzucato2016stimuli,engel2016selective}. In all those cases, and consistent with our results, state dwell times showed large trial-to-trial variability captured by Markovian dynamics (i.e. right-skewed distributions), suggesting an underlying stochastic process driving transitions \cite{MillerKatz2010,mazzucato2015dynamics,mazzucato2019expectation}. A novel feature of our results is that the sequence of attractors is not driven by external stimuli, but rather is internally generated.

\subsection{Network models of sequences}

The main features of M2 ensemble activity targeted by our two-area mesoscale attractor network (2AMAN) model were the reliable identity and order of long-lived attractors occurring in a sequence, and the large trial-to-trial variability of attractor dwell times, whose distributions are characterized by a positive skewness and large CV. In our model, we showed that both features can be robustly attained when transitions between attractors arise from correlated variability. 

Previous network models could achieve either sequence reliability or variability in dwell time distributions, but not both. Models generating reliable pattern sequences include synfire chains \cite{abeles91,diesmann99}. These models rely on a fine tuned connectivity structure producing pattern dwell times with short duration and low variability. While these dynamics are well suited to explain neural activity observed in songbird HVC  \cite{hahnloser2002ultra,fiete2010spike} or mammalian hippocampus \cite{nadasdy1999replay}, their features are not compatible with the observed M2 ensemble activity. Reliable pattern sequences can otherwise be triggered by specific cues in recurrent networks with asymmetric connectivity structure \cite{sompolinsky1986temporal,kleinfeld1986sequential,dehaene1987neural,treves2005frontal,murray2017learning,gillett2019characteristics}, trained with unsupervised learning rules \cite{jun2007development,liu2009embedding, fiete2010spike, pereira2018unsupervised} or in reservoir networks \cite{rajan2016recurrent}. However, pattern dwell times in such models are short, set by single-neuron characteristic time constants, and show little trial-to-trial variability. Pattern dwell times could be increased via synaptic delays \cite{sompolinsky1986temporal}. However, none of these models is capable of generating large trial-to-trial variability in dwell time distributions and are thus incompatible with the observed M2 data.

Long-lived patterns of neural activity can be sustained by attractor dynamics, where the reverberating activity of neural assemblies is sculpted by the recurrent couplings \cite{hopfield1982neural,amit1985spin, Amit1997a}. Previous attractor networks were shown to generate sequences of long-lived metastable patterns whose features, however, are incompatible with the ones we observed in M2 neural ensembles. In particular, networks with clustered architecture can give rise to metastable attractors with large trial-to-trial variability in dwell time distributions \cite{LitwinKumarDoiron2012,DecoHugues2012,mazzucato2015dynamics,mazzucato2019expectation}. However, metastable attractors in these models concatenated in random sequences, incompatible with the highly reliable sequences we observed in M2. The reason is that, in these networks, each cluster generates independent fluctuations within activity space, realizing a high-dimensional stochastic process, akin to the private noise model in \Cref{fig:S7}. These fluctuations drive transitions along random directions in activity space, thus unreliable across trials (when concatenating more than two states in a sequence \cite{mongillo2003retrospective}). To drive a specific transition, independent fluctuations would have to align along a specific direction within the high-dimensional activity space, and the probability of this event occurring vanishes for large network size. Thus it is challenging to generate reliable yet noise-driven sequences in these models. We confirmed this intuition by showing that no regime of parameters allowed transitions with right-skewed attractor dwell times in a private noise model (\Cref{fig:S7}). One may drive clustered networks with strong time-dependent stimuli to pace activity along stimulus-specific sequences  \cite{MillerKatz2010,mazzucato2015dynamics,mazzucato2016stimuli,mazzucato2019expectation,darshan2017canonical}. However, this would merely shift the problem of reliable sequence generation from the local circuit to an upstream area producing the specific input (but see \cite{bernstein2017markov}). 

By introducing correlated variability to attractor networks, our 2AMAN model provides a circuit mechanism to overcome the curse of dimensionality. By constraining noise correlations onto a low-dimensional manifold, the 2AMAN model attained reliable sequences of long-lived attractors with large trial-to-trial variability in dwell time distributions.

\subsection{Neural circuits underlying attractor dynamics}

Our 2AMAN mechanistic model naturally captured the essential features of the observed M2 population dynamics. How does the model architecture map onto specific neural circuits? Previous work showed, using inactivation experiments, that the stochastic component in action timing variability originated downstream of the medial prefrontal cortex and presumably in M2 \cite{murakami2017distinct}. Our model provides a mechanistic explanation in the form of correlated variability, inherited from synaptic couplings between a large recurrent network representing M2 and a smaller and faster network lacking recurrent couplings. We tentatively hypothesize that this latter network is instantiated by a small subcortical circuit connected to M2, such as the areas that comprise its basal ganglia or thalamic partners. Recent evidence from perturbation experiments in rodents supports a scenario where a mesoscale circuit sustaining attractor dynamics includes the thalamus and motor \cite{guo2017maintenance} or prefrontal cortex \cite{schmitt2017thalamic}. This hypothesis could be directly tested via perturbation experiments, and we leave this question for future work. It is plausible that a larger mesoscale network may underlie sequence generation, including cortex, thalamus, and basal ganglia \cite{kawai2015motor,helie2015learning,desmurget2010motor,nakajima2019prefrontal}. Even though our results suggest that timing variability of multi-step sequences extending on the order of 10 seconds may originate within a cortical-subcortical loop, we did not investigate how even more complex natural behavior could emerge at longer timescales \cite{wiltschko2015mapping,berman2016predictability}. This may also involve a larger and more distributed mesoscale network \cite{svoboda2018neural}. Flexibly switching between different action sequences to adapt to a changing environment may recruit other areas including the anterior cingulate \cite{murakami2017distinct} and the basal ganglia \cite{markowitz2018striatum,jin2015shaping,murray2017learning,nakajima2019prefrontal,kao2005contributions}. We hope to address these broader issues in future work.

A large amount of evidence implicated preparatory activity in rodent M2, specifically the antero-lateral motor cortex, in action and movement planning both in forced-choice tasks \cite{erlich2011cortical,li2015motor,chen2017map,sul2011role,inagaki2019discrete, guo2014flow} as well as self-initiated tasks \cite{murakami2014neural,murakami2017distinct}. The sequences of metastable attractors we uncovered in M2 were consistent with the features of preparatory activity \cite{jin2015shaping}: a precise dictionary linked specific attractors to actions; attractor onset reliably predicted action onset; and action timing variability strongly correlated with attractor onset variability. In particular, we were able to predict actions in incorrect trials using the cross-validated attractor/action dictionary established in correct trials, consistent with recent findings \cite{wei2019orderly}. This interpretation is further supported by the fact that M2 population activity was only modulated by self-initiated actions and M2 neurons were not responsive to other task features such as trial type (patient vs. impatient, or correct vs. incorrect) and reward expectation (large vs. short).

\subsection{Correlated variability in sensory vs. motor processing}

The main conceptual innovation in our 2AMAN model is the introduction of low-dimensional correlated variability driving reliable sequences with variable timing. Similar ``motor noise correlations" have been recently reported during vocal babbling in juvenile songbirds. Correlated variability was present in a motor area (RA) but absent in a premotor area (LMAN) \cite{darshan2017canonical}, and a mechanism for the emergence of correlated variability via topographic organization of projections from LMAN to RA was proposed.  This is in contrast to our observations of correlated variability in M2, a premotor area. In our 2AMAN model, correlated variability arises from a feedback loop between a high- and a low-dimensional recurrent network, mediated by asymmetric couplings in the synaptic connectivity matrix. Asymmetric couplings  have been previously used to generate specific temporal dynamics, though in the absence of noise. Examples include models of temporal sequences \cite{sompolinsky1986temporal,kleinfeld1986sequential} or recurrent networks within the echo-state/reservoir computing framework \cite{maass2002real,jaeger2004harnessing,sussillo2009generating,mastrogiuseppe2018linking}. All these models are fundamentally different from ours, as their low-rank structure is fixed and time-independent, hence their temporal dynamics entails no trial-to-trial variability. In our model, on the other hand, the low-rank structure is time-varying and generates correlated fluctuations at a fast timescale. We confirmed that low-dimensional correlated fluctuations around each attractor are present in the empirical M2 data. Alternative models where each neuron's fluctuations are independent (i.e., with private noise) failed to capture the observed temporal variability and predicted high-dimensional fluctuations, which were absent in the empirical data.

Low-dimensional correlated variability has been widely reported in sensory cortex, where it may carry information about the animal's state of locomotion \cite{niell2010modulation} or arousal \cite{polack2013cellular,mcginley2015waking}, facial and whisker movements  \cite{stringer2018spontaneous,musall2019single,salkoff2019movement}, or attentional state \cite{cohen2009attention,huang2019circuit}. It has been proposed that low-dimensional correlated variability in sensory cortex, in the form of differential correlations, may be detrimental to sensory processing as it may limit a network's information processing capability \cite{moreno2014information}. Alternatively, correlated variability may arise from top-down feedback projections carrying task-related information \cite{bondy2018feedback}. Here, we found that low-dimensional correlated fluctuations are the crucial mechanism enabling neuronal sequences to unfold with variable timing. It is likely that variable timing is an adaptive feature of motor behavior. Amongst other possible functions, such as avoiding predation or competition, timing variability allows animals to explore the temporal aspects of a given sequence of behavior independently of the choices of actions. We speculate that exploration could allow learning of proper timing by a search in timing space independent of action selection and vice-versa, as may be the case in songbirds \cite{kao2005contributions,goldberg2011vocal,darshan2017canonical}. Our results thus suggest that low-dimensional correlations are essential for motor generation.
\FloatBarrier

\newcommand{\beginsupplement}{%
        \setcounter{table}{0}
        \renewcommand{\thetable}{S\arabic{table}}%
        \setcounter{figure}{0}
        \renewcommand{\thefigure}{S\arabic{figure}}%
     }
\beginsupplement
\begin{figure*}[th!]
\captionsetup{justification=raggedright}
\subfloat{\label{fig:S1a}}
\subfloat{\label{fig:S1b}}
\subfloat{\label{fig:S1c}}
\subfloat{\label{fig:S1d}}
\centering
\includegraphics[width=1.\textwidth]{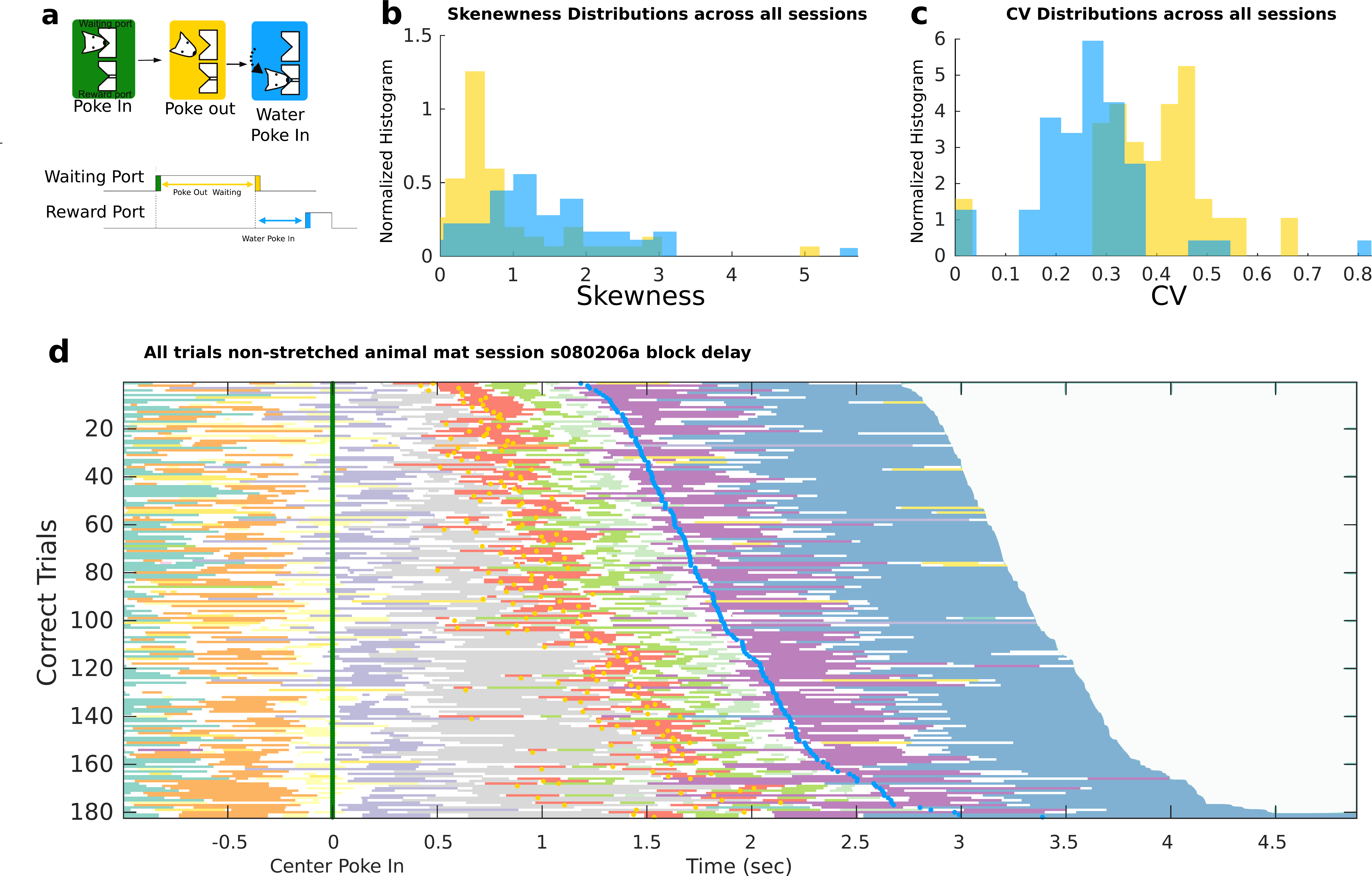}
\centering
\caption{{\bf Interevent interval distributions}. \protect\subref{fig:S1a}) Schematic of the interevent time intervals between Poke In and Poke Out (yellow) and between Poke Out and Water Poke In (blue). \protect\subref{fig:S1b}) Histogram of skewness of interevent interval distributions across all sessions revealed right-skewed distributions. \protect\subref{fig:S1c}) Histogram of coefficients of variability (CV) of interevent intervals across all sessions revealed large trial-to-trial variability variability in interevent times. \protect\subref{fig:S1d}) Pattern sequences in all trials, ordered by duration (blue dots represent Water Poke In). As opposed to \protect\Cref{fig:1e} where trials time courses were stretched, trials in \protect\subref{fig:S1d}) represent actual time courses.}
\label{fig:S1}
\end{figure*}

\begin{figure*}[th!]
\captionsetup{justification=raggedright}
\subfloat{\label{fig:S2a}}
\subfloat{\label{fig:S2b}}
\subfloat{\label{fig:S2c}}
\subfloat{\label{fig:S2d}}
\centering
\includegraphics[width=1\textwidth]{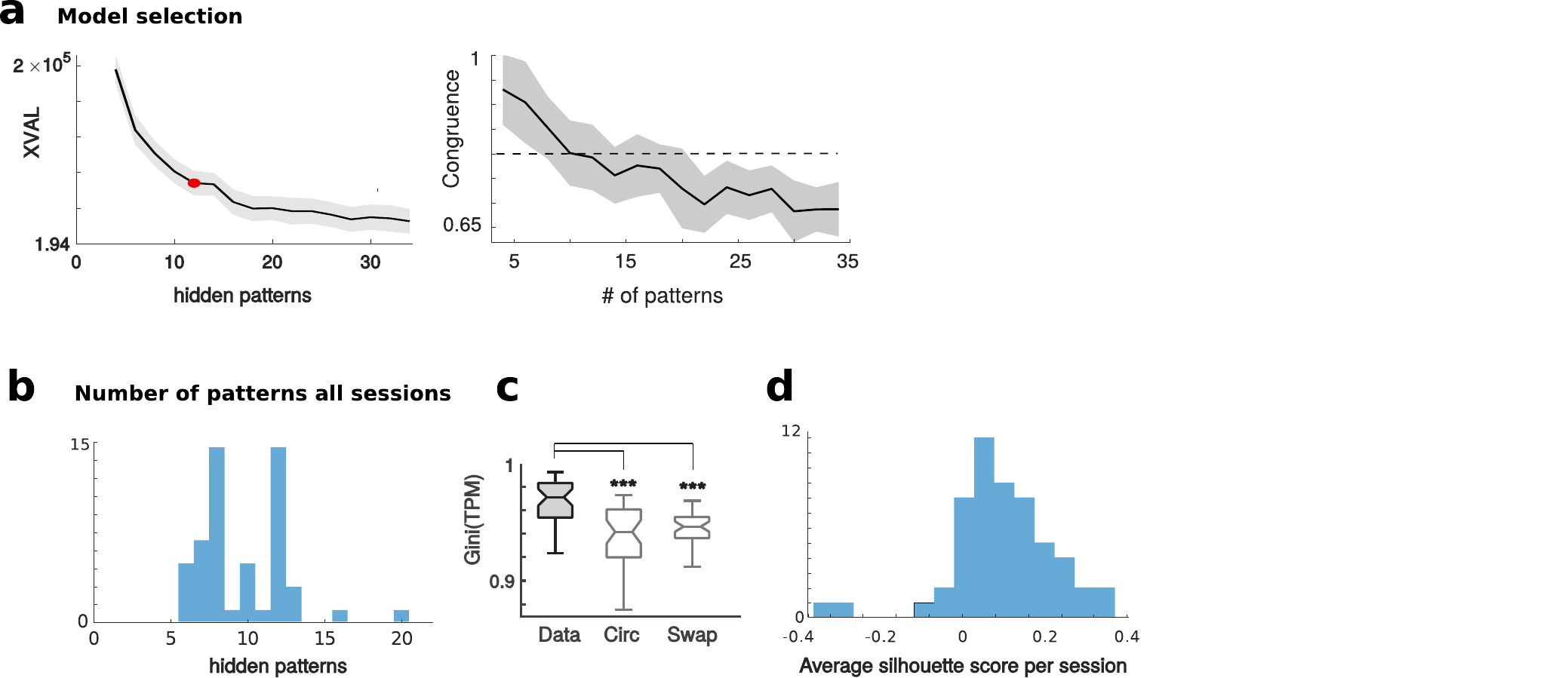}
\centering
\caption{{\bf Empirical data fit via hidden Markov models (HMM).} \protect\subref{fig:S2a}) Model selection in a representative session. Left: The number of patterns was selected via an unsupervised 10-fold cross-validation procedure. The number of patterns yielding the largest curvature in the likelihood trend is selected (XVAL=$-2\times $log-likelihood of held-out trials plateaus for increasing number of patterns). Right: For a fixed number of patterns, the similarity between the HMM parameters (cross-validated congruence, see Methods) optimized in different folds drops below 0.8 (dashed line) when the number of patterns grows beyond the value selected via cross-validation (see \cite{williams2018unsupervised,tomasi2006comparison}). \protect\subref{fig:S2b}) Distribution of the number of patterns across all sessions. \protect\subref{fig:S2c}) Gini coefficient distribution for the pattern transition probability matrices (TPM) across all sessions, compared to shuffled datasets. Empirical TPMs are sparser than TPMs inferred from surrogate datasets (see \Cref{fig:2}, $***=p<0.001$). \protect\subref{fig:S2d}) Distribution of within- and across-cluster distances between patterns measured with silhouette scores (ranksum test, $p<2.0\times10^{-7}$, see \Cref{fig:3a}). }
\label{fig:S2}
\end{figure*}

\begin{figure*}[th!]
\captionsetup{justification=raggedright}
\subfloat{\label{fig:S3a}}
\centering
\includegraphics[width=1.\textwidth]{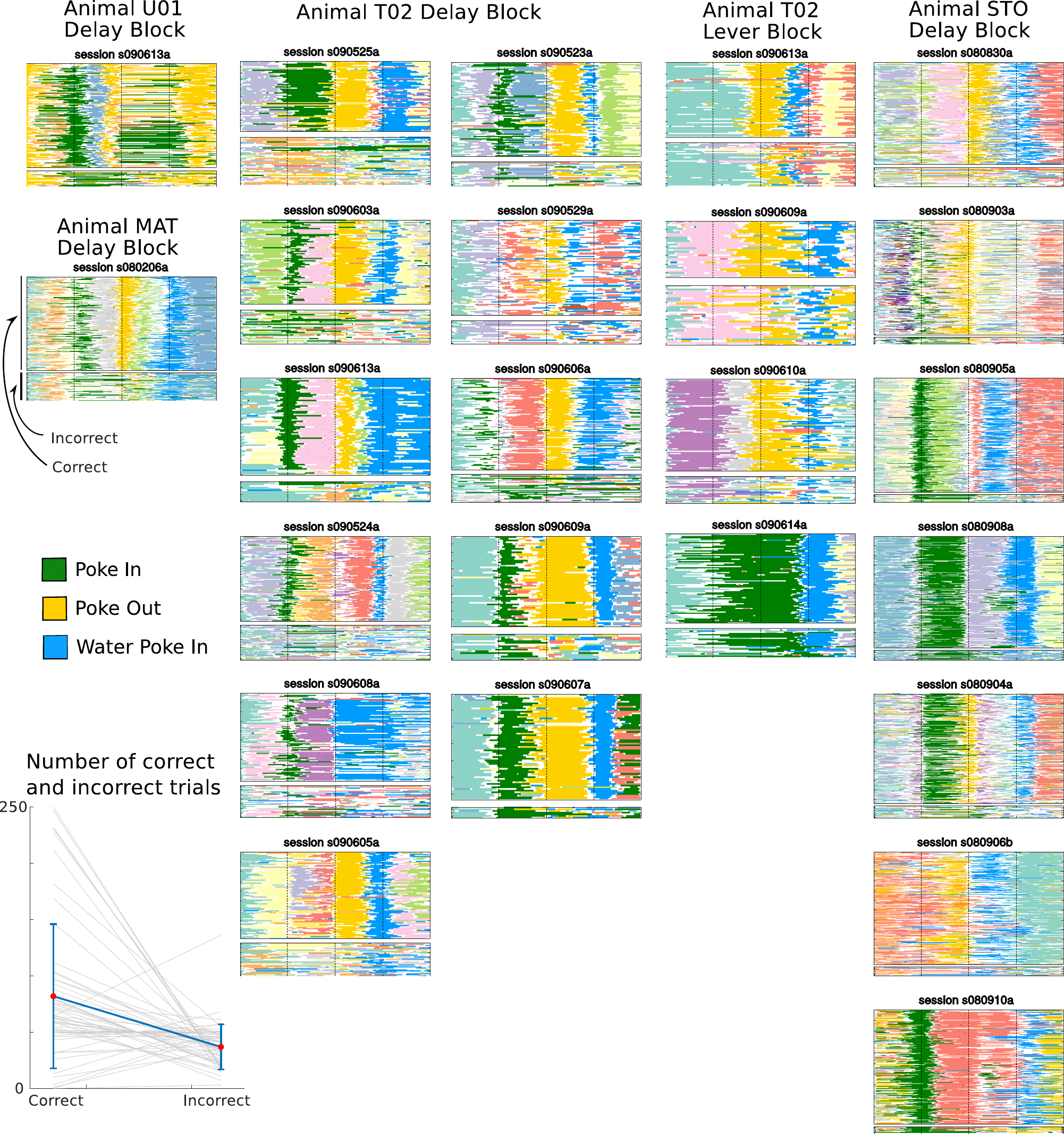}
\centering
\caption{{\bf Stretched and event-aligned pattern sequences for all sessions.} In each session, neural patterns from correct trials (top subplots, same notations as in \Cref{fig:1e}) show reliable sequences, as expected from the fixed action sequences to be performed to collect a reward; pattern sequences in incorrect trials are less reliable (bottom subplots, see \Cref{fig:4b}), as expected from the inconsistent behavior in those trials. In most sessions, animals only performed the delay task (``Delay block''); in sessions where delay and lever tasks (``Lever block'') were interleaved, block trials from the two tasks were plotted separately (only for animal T02). The patterns tagged to one of the three events analyzed (cf. legend and \Cref{fig:4}) are consistently colored across session. The remaining patterns do not follow a consistent color code. Inset in bottom left: Number of correct and incorrect trials in each session. Session s090206a is the representative session in \Cref{fig:1}.}
\label{fig:S3}
\end{figure*}

\begin{figure*}[th!]
\captionsetup{justification=raggedright}
\subfloat{\label{fig:S4a}}
\centering
\includegraphics[width=1\textwidth]{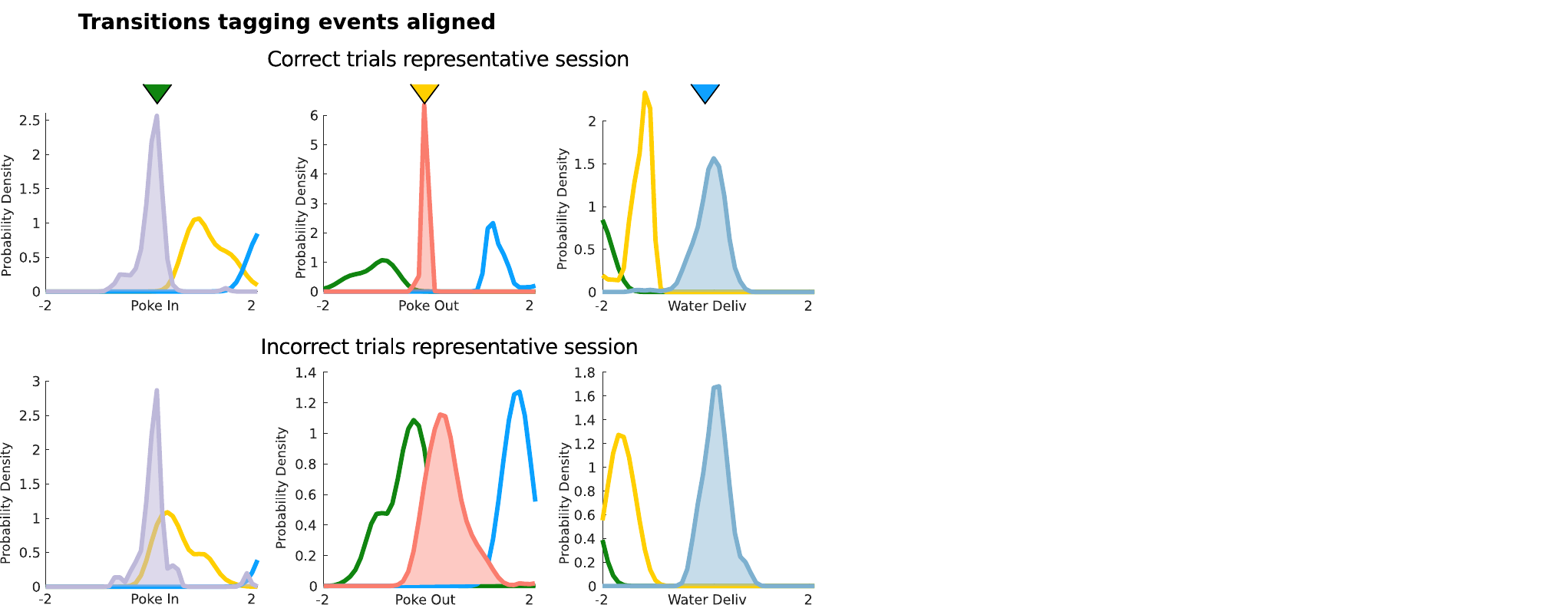}
\centering
\caption{{\bf Predicting actions from patterns in incorrect trials}. After tagging transitions to events using correct trials (top row, left to right: Poke In, Poke Out, Water Poke In; see \Cref{fig:4}), transitions were subsequently aligned to the same events but in incorrect trials (bottom row). In each subplot, the distribution of transition times for the corresponding event (e.g. Poke In in the top left: purple-filled histogram shows distribution of ``Poke In'' transition times aligned to the Poke In events) is compared to events time distribution for other events (yellow and blue curves).}
\label{fig:S4}
\end{figure*}

\begin{figure*}[th!]
\captionsetup{justification=raggedright}
\subfloat{\label{fig:S5a}}
\subfloat{\label{fig:S5b}}
\centering
\includegraphics[width=1.\textwidth]{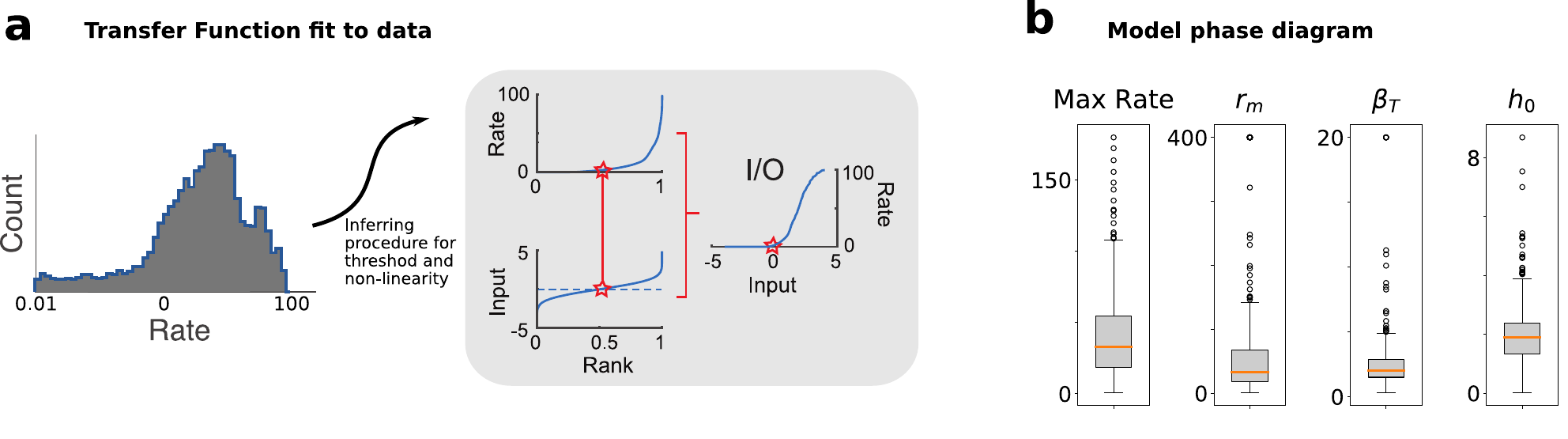}
\centering
\caption{{\bf Inferring transfer functions from pattern sequences} \protect\subref{fig:S5a}) Procedure to infer single-neuron current-to-rate transfer functions from the data. The empirical distribution of firing rates across patterns for a representative neuron (left) was rank-matched to a standardized normal distribution of input currents (top and bottom left panels in grey box), obtaining the current-to-rate function (right).  The star in each plot corresponds to the median value. \protect\subref{fig:S5b}) Each single-cell current-to-rate function was fit to a sigmoidal function, yielding a distribution of fit parameters (366 neurons from 41 sessions; see Methods, Eq.~(\ref{transfer:1})).}
\label{fig:S5}
\end{figure*}

\begin{figure*}[th!]
\captionsetup{justification=raggedright}
\subfloat{\label{fig:S6a}}
\subfloat{\label{fig:S6b}}
\centering
\includegraphics[width=0.8\textwidth]{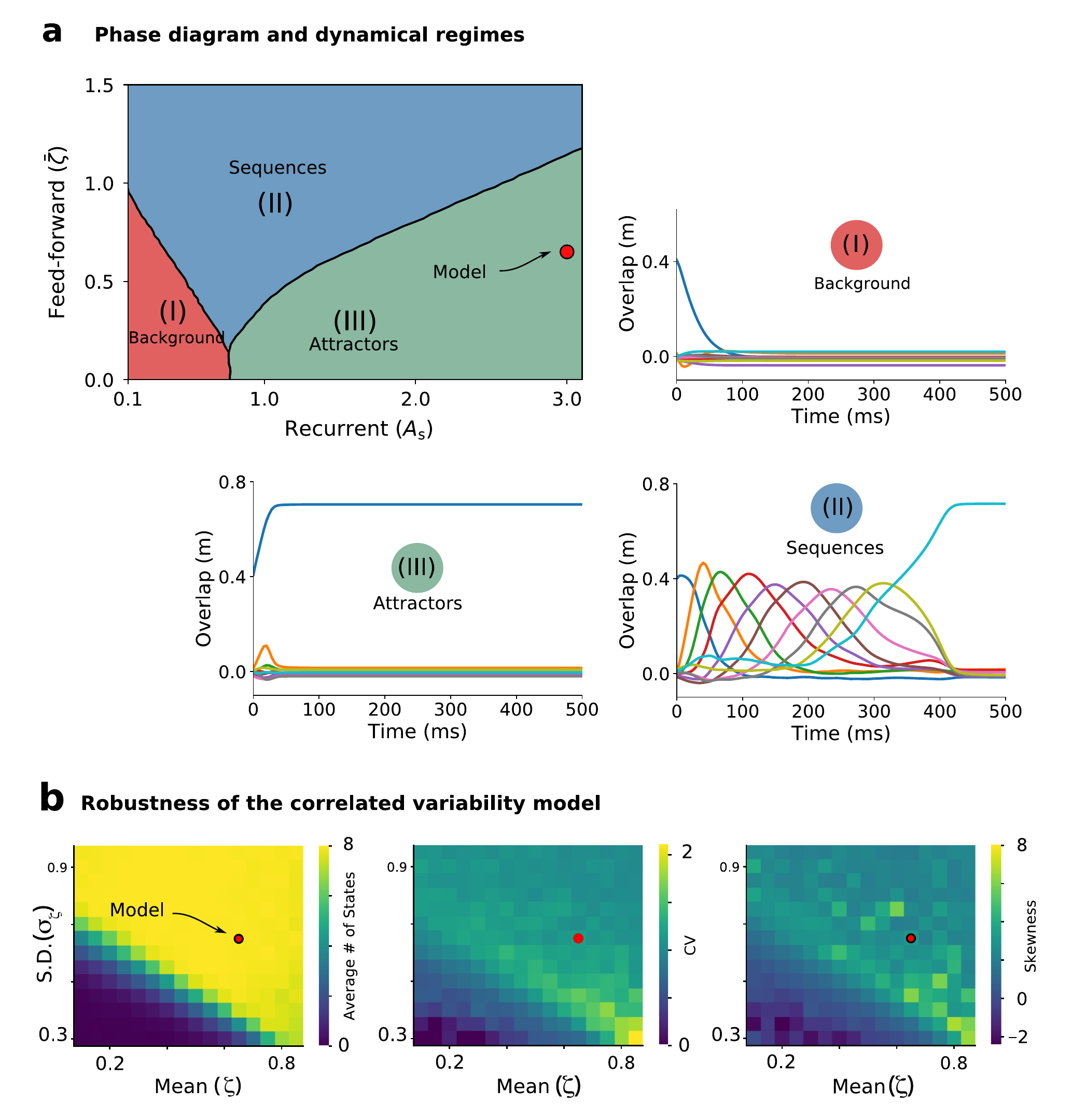}
\centering
\caption{{\bf Phase structure and model robustness.}
\protect\subref{fig:S6a}) Phase diagram for a recurrent network in the absence of noise (model in Eq.~(\ref{model:1}) with $\sigma_\zeta=0$; average over 100 network realizations). The number of attractors visited per trial depends on the strength of the average recurrent (x-axis: the parameter $A_S$ represent the strength of $J^S$ in Eq.~(\ref{model:2})) and feedforward couplings (y-axis: the parameter $\bar\zeta$ represents the strength of $J^F$ in Eq.~(\ref{model:2})). The model activity decays to zero for low values of the couplings (phase I, red); it generates stable attractors for large recurrent weights (phase II, green) or sequences of attractors for large feedforward weights (phase III, blue). In the representative trials, color-coded curves represent the time course of the overlaps (see Eq.~(\ref{model:overlap})) between network activity and attractors. The red dot in the phase diagram shows the values of $J^S$ and $J^F$ used in the Results for the correlated variability model (upon adding multiplicative noise for $J^F$). \protect\subref{fig:S6b}) Robustness of the correlated variability model. When sistematically varying the mean $\bar \zeta$ and the standard deviation $\sigma_\zeta$ of the noise $\zeta(t)$ in correlated variability model (corresponding to the value of $J^S$ and $J^F$ from panel \protect\subref{fig:S6a}), red dot), the network robustly generates long attractor sequences (with an average of $p\approx 8$ attractors in each sequence, left) with large CV (center) and skewness (right) in attractor dwell time distributions over 200 trials.  
}
\label{fig:S6}
\end{figure*}

\begin{figure*}[th!]
\captionsetup{justification=raggedright}
\subfloat{\label{fig:S7a}}
\subfloat{\label{fig:S7b}}
\centering
\includegraphics[width=0.9\textwidth]{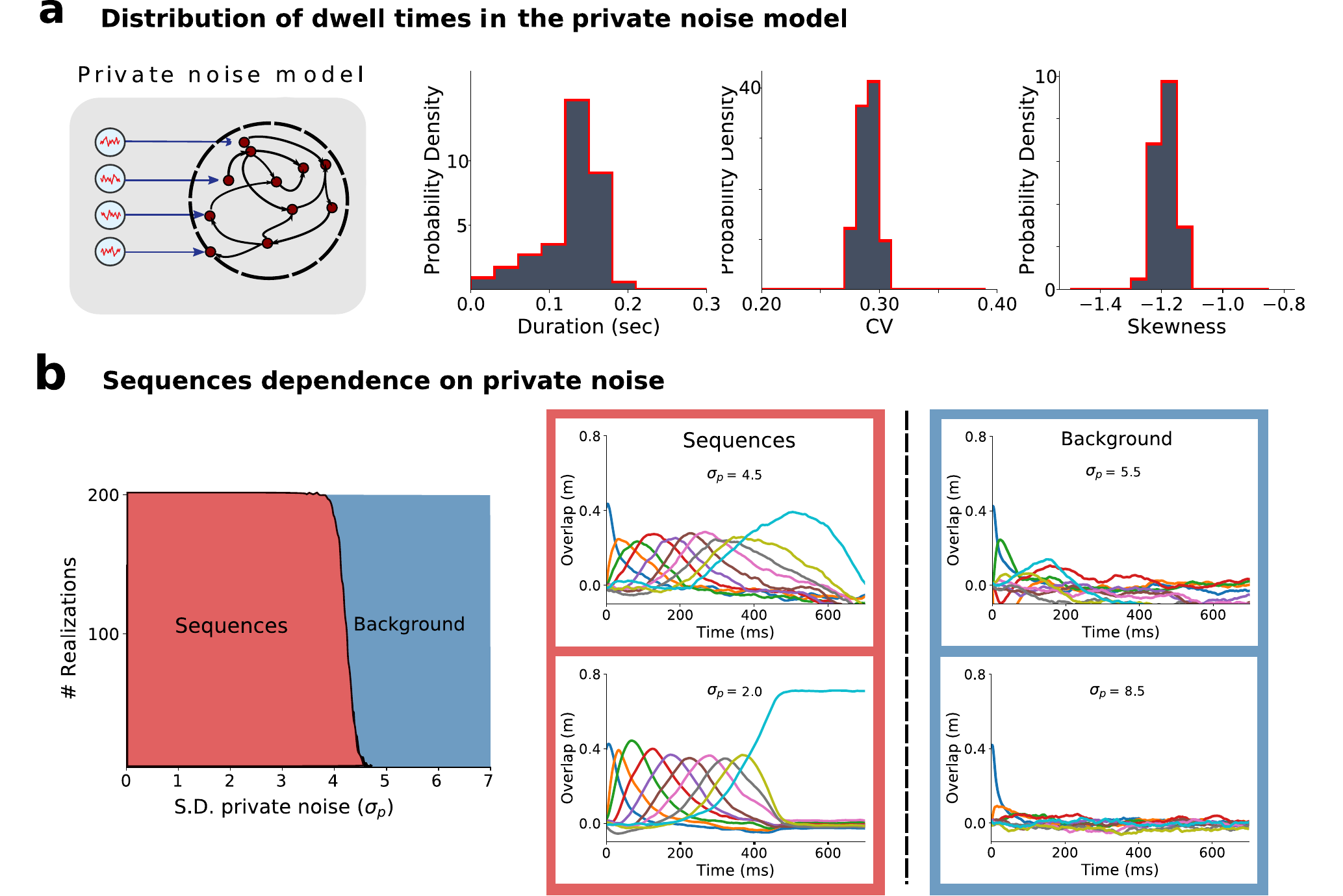}
\centering
\caption{{\bf Phase structure of models with recurrent and feedforward couplings.}
\protect\subref{fig:S7a}) In a model with recurrent and feedforward couplings, adding private noise with variance (see Methods, Eq.~(\ref{model:privatenoise}; network parameters as in Table \ref{table:1}, except for $A_S=1$, $\bar \zeta=0.65$, $\sigma_\zeta=0$, $\sigma_p=2$) introduces a small amount of trial-to-trial variability in pattern dwell time distributions (left), yielding small non-zero CVs (center) and negative skewness (right) across different networks. The dwell time statistics are qualitatively different from the empirical ones (cf. \Cref{fig:1f}). \protect\subref{fig:S7b}) Left: Beyond a critical value of noise strength $\sigma_p$, the private noise model breaks down. For low private noise, networks generate sequential dynamics (red area, networks with sequential dynamics out of 200 network realizations), though with low CV and skewness; for larger values of private noise, activity decays to zero (blue). Right: Representative trials with 4 different values of the noise strength $\sigma_p$.
}
\label{fig:S7}
\end{figure*}

\section{Methods}
\subsection{Experimental procedures} 
\noindent{\it Behavioral task}. Rats were trained on the self-initiated waiting task (\Cref{fig:1a}) in a behavioral box containing a Wait port at the center and a Reward port at the side (entry/ exit from ports were detected via infrared photo-beam). Rats self-initiated a trial by poking into the Wait port (``PokeIn''). If the rat stayed in the Wait port for T1 delay (0.4s), the first tone played (tone 1; 6 or 14 kHz tone), signaling availability of reward in the Reward port. If the rat waited in the Wait port after tone 1, then tone 2 was played after a T2 delay (14 or 6 kHz, different from tone 1). If the rat visited the Reward port after tone 2, a large water reward (40 $\mu$l) was delivered after a 0.5s delay (patient trial). If the rat poked out after tone 1 but before tone 2, and visited the Reward port, a small water reward (10 $\mu$l) was delivered after a 0.5s delay (impatient trial). The rat had to visit the Reward port within 2s after the poke out to collect rewards. These trials were referred to as ``correct trials;'' trials were the animal performed different action sequences were deemed ``incorrect trials.'' If the rat poked out before tone 1, no rewards were made available. Re-entrance to the Wait port was discourage with a brief noise burst. T2 delay was drawn from an exponential distribution, with minimum value 0.7s and mean adjusted to achieve patient trials in one third of the session. After reward delivery, an inter-trial interval (ITI) started during which white noise played. The time from the PokeIn to the ITI end was held constant, so that the rat could not increase reward collection by leaving the Wait port fast with the goal to start the next trial early. The optimal strategy was thus to always wait for tone 2. To test whether neuronal responses depended on a specific action, 3 rats were trained on two variants of the task. In these experiments, a different behavioral box contained a Reward port, a nose-poke Wait port, and a lever-press Wait port. Blocks of nose-poke trials and lever-press trials were interleaved in each session. In the nose-poke block, the rat was to perform the same task as above. In the lever-press block, task rules were the same but the rat had to wait for the tones by keeping the lever pressed. The wrong action (nose-poke waiting in the lever-press block and vice versa) was not rewarded and classified as ``incorrect
trials.'' Each block last for 70-100 trials. Transitions between the blocks were not signaled. 41 sessions (7 rats) were recorded, see \cite{murakami2014neural} for extensive details. 

\noindent{\it Electrophysiological data}. Rats were implanted with a drive containing 10-24 movable tetrodes targeted to the M2 (3.2-4.7 mm anterior to and 1.5-2.0 mm lateral to Bregma). Electrical signals were amplified and recorded using the NSpike data acquisition system (L.M. Frank, University of California, San Francisco, and J. MacArthur, Harvard University Electronic Instrument Design Lab). Multiple single units were isolated offline by manually clustering spike features derived from the waveforms of recorded putative units using MCLUST software (A.D. Redish, University of Minnesota). Tetrode depths were adjusted before or after each recording session in order to sample an independent population of neurons across sessions. See \cite{murakami2014neural} for details.

\subsection{Neural data analysis} Data analyses was performed with custom-written software using MATLAB (Mathworks) and Python. No statistical methods were used to pre-determine sample sizes, but sample sizes were similar to previous studies \cite{erlich2011cortical,guo2014flow}. All summary statistics are mean$\pm$SD across 41 sessions, unless otherwise stated. 

\subsection{Pattern sequence estimation}
A Poisson-Hidden Markov Model (HMM) analysis was used to detect neural pattern sequences from simultaneously recorded activity of ensemble neurons. Here, we briefly describe the method used and refer to Refs.\cite{mazzucato2015dynamics,mazzucato2019expectation} for details.
According to the HMM, the network activity is in one of $M$ hidden ``patterns'' at each given time. A pattern is a firing rate vector $r_i(m)$ (the ``emission matrix'', \Cref{fig:1c}), where $i= 1,\ldots,N$ is the neuron index and $m=1,\ldots,M$ identifies the pattern. In each pattern, neurons discharge as stationary Poisson processes (Poisson-HMM) conditional on the pattern's firing rates $r_i(m)$.  Stochastic transitions between patterns occur according to a Markov chain with transition matrix (TPM, \ \Cref{fig:1c}) $T_{mn}$, whose elements represent the probability of transitioning from pattern $m$ to $n$ at each given time. We segmented trials in 5 ms bins, and the observation of either $y_i(t)=1$ (spike) or $y_i(t)=0$ (no spike) was assigned to a bin at time $t$ for the $i$-th neuron (Bernoulli approximation); if in a given bin more than one neuron fired, a single spike was randomly assigned to one of the active neurons. A single HMM was fit to all correct trials per  session, yielding emission probilities and transition probabilities between patterns, optimized via the Baum-Welch algorithm with a fixed number of hidden patterns $M$ (iterative maximum likelihood estimate of parameters and latent patterns given the observed spike trains). 

The number of patterns $M$ is a model hyperparameter, optimized using the following model selection procedure \cite{engel2016selective}. In each session, we used K-fold cross-validation (with $K=20$) to train an HMM on $(K-1)-$folds and estimate the log-likelihood of the held-out trials $LL(M)$ as a function of number of patterns $M$ in the fit (see \Cref{fig:S2}). The held-out $LL(M)$ increases with $M$, until reaching a plateau. We selected the number of patterns $M^*$ for which the incremental increase $LL(M+1)-LL(M)$ had the largest drop (the point of largest curvature) before the plateau. For control, we performed model selection using an alternative method, the Bayesian Information Criterion  \cite{mazzucato2019expectation}, obtaining comparable results (not shown). 

To gain further insight into the structure of the model selection algorithm, we performed a post-hoc comparison between the parameters optimized on the training set for each value of $M$ (number of patterns), across the cross-validation K-folds. In particular, we estimated the similarity between the optimized features (emission $r^{[k_1]}_i(m)$ and transition matrices $T^{[k_1]}_{mn}$) in the $k_1$-th fold and the $k_2$-th fold for given $M$, according to the following congruence $C(k_1,k_2)$ measure \cite{tomasi2006comparison}:
$$
C(k_1,k_2)=\left(\sum_{m=1}^M \sum_{i=1}^N \hat r^{[k_1]}_i(m)\hat r^{[k_2]}_i(m)\right)\cdot\left(\sum_{m,n=1}^M\hat T^{[k_1]}_{mn}\hat T^{[k_2]}_{mn}\right) \ ,
$$
where $N$ is the ensemble size, $\hat r^{[k]}_i(m)=r^{[k]}_i(m)/||\overrightarrow{r}^{[k]}(m)||_2$ is the normalized emission for pattern $m$, and $\hat T^{[k]}_{mn}$ is the normalized transition matrix $\hat T^{[k]}_{mn}=T^{[k]}_{mn}/||T^{[k]}||_2$. Features were matched across folds using the stable matching algorithm \cite{gale2013college}.
If the two folds yielded identical parameters, one would find $C(k_1,k_2)=1$. A congruence above 0.8 signals good quantitative agreement between different folds, whereas congruence below 0.6 suggests a poor similarity among folds \cite{williams2018unsupervised}. We calculated the average congruence across all fold pairs for given $M$ and verified that the number $M^*$ of patterns selected with the cross-validation procedure above corresponded to the elbow in the congruence curve (see \Cref{fig:S2a}). For larger number of patterns, average congruence typically fell below 0.8.

The Baum-Welch algorithm only guarantees reaching a local rather than global maximum of the likelihood. Hence, for each session, after selecting the number of pattern $M^*$ as above, we ran 20 independent HMM fits on the whole session, with random initial guesses for emission and transition probabilities, and kept the best fit for all subsequent analyses. The winning HMM model was used to infer the posterior probabilities of the patterns at each given time $p(m,t)$ from the data. Only those patterns with probability exceeding $80\%$ in at least 50 consecutive ms were retained (henceforth denoted simply as patterns, \Cref{fig:1d}). This procedure eliminates patterns that appear only very transiently and with low probability, also reducing the chance of over-fitting. Pattern dwell time distributions (\Cref{fig:1f}) within each session were estimated from the empirical distribution of interval times where a pattern's probability was above $80\%$. 

\subsection{\label{methods:comparison}Comparison with surrogate datasets} 
We compared the HMM analysis of the empirical dataset with two surrogate datasets, obtained with the following shuffled procedure (\Cref{fig:2}, \cite{maboudi2018uncovering}). In the ``circular'' shuffle, each neuron's binned spike counts were circularly shifted within-trial randomly (row-wise circular shift), preserving autocorrelations but destroying pairwise correlations. In the ``swap'' shuffle, binned population spike counts were randomly permuted in time (column-wise swap), preserving pairwise correlations but destroying autocorrelations. For comparison of the real dataset with shuffled ones, we adopted the same K-fold cross-validation procedure as above, where an HMM was fit on training sets and the posterior probabilities $p(m,t)$ of patterns were inferred from observations in the held-out trials (test set).

From the pattern posterior probabilities inferred on held-outs, we estimated several observables for comparison between real and shuffled datasets. Pattern detection confidence was estimated as the fraction of a trial length where a pattern was detected with high confidence ($p(m,t)>80\%$). Sparseness of transitions was estimated as the average Gini coefficient of TPMs obtained from the K training sets. We also estimated the across-trials sequence similarity as follows. In a trial where patterns were detected above $80\%$ in a certain consecutive order, we compiled a ``symbolic'' TPM, whose diagonal element $T^{(sym)}_{mm}$ were set equal to the number of non-consecutive occurrences of pattern $m$, and off-diagonal element $T^{(sym)}_{mn}$ was set equal to the number of $n\to m$ transitions observed; finally each row was normalized: $T^{(sym)}_{mn}\to T^{(sym)}_{mn}/\sum_{l=1}^N T_{ml}$. E.g. the pattern sequence $1,2,3,1,2$ is in one-to-one correspondence to the symbolic TPM 
$$
\textrm{sequence \, [1,2,3,1]}\,\leftrightarrow\, T^{(sym)}_{mn}=\left(\begin{array}{ccc}0.67 &0.33& 0\\
0 & 0.5 &0.5\\
0.5& 0 &0.5\end{array}\right) \ .
$$
Sequence similarity was defined as the trial-averaged Pearson correlation between $T^{(sym)}$.

In the data, we define the overlaps $q$ between $N$-dimensional vectors $r_i$ and $s_i$ describing inferred patterns as the correlation coefficient
$$
q[r,s]={1\over N}\sum_{i=1}^N {r^is^i\over\sigma(r)\sigma(s)} \ ,
$$
where $\sigma(r)$ is the standard deviation of $r^i$.

\subsection{\label{methods:multistability}Single neuron multistability}

To assess how single-neuron activity was modulated across different patterns, local (i.e., single-trial) firing rate estimates for neuron $i$ given a pattern $m$ were obtained from the maximization step of the Baum-Welch algorithm
\begin{equation}
\label{eq:hmmrate}
r_i(m)=-{1\over dt}\log\left(1-{\sum_{t=1}^T p(m,t)y_i(t)\over \sum_{t=1}^T p(m,t)}\right) \ ,
\end{equation}
where $y_i(t)$ are the neuron's observations in the current trial of length $T$. To determine whether a neuron's conditional firing rate distributions differed across patterns (\Cref{fig:3c}), we performed a non-parametric one-way ANOVA (unbalanced Kruskal-Wallis, p<0.05). A post-hoc multiple-comparison rank analysis (with Bonferroni correction) revealed the smallest number of significantly different firing rate distributions across patterns. Given a p value $p_{mn}$ for the pairwise post hoc comparison between patterns $m$ and $n$, we considered the symmetric $M\times M$ matrix $S$ with elements $S_{mn}=0$ if the rates were different ($p_{mn}<0.05$) and $S_{mn}=1$ otherwise. For example, consider the case of 3 patterns and the following A matrix, where patterns were sorted by firing rates:
$$ 
S=\left(
\begin{array}{ccc}
\cdot &1 &0 \\
\cdot &\cdot& 0 \\
\cdot &\cdot& \cdot
\end{array}
\right) \ .
$$
Firing rates of patterns 1 and 2 were not significantly different, but they were different from pattern 3 firing rate. Hence, in this case we classified the neurons as multistable with 2 different firing rates across patterns \cite{mazzucato2015dynamics}.

\subsection{Tagging pattern onsets to self-initiated actions}
The HMM analysis yields a posterior probability distribution $p(m,t)$ for the neural pattern $m$ at time $t$. At any time $\bar{t}$ we identified the active pattern $\bar{m}$ when $p(\bar{m},\bar{t})\ge 0.8$. In case this criterion was not met by any pattern then no pattern was assigned, cfr \Cref{fig:1d}. The onset time of a specific pattern $\bar{m}$ was identified as the first time $\bar{t}$ where $p(\bar{m},\bar{t})\ge 0.8$. Transitions of several patterns appeared in close proximity to specific events \Cref{fig:4a}, we thus developed a method to tag pattern onsets to specific events. Specifically we tagged onset of a given pattern with one of three actions (Poke In, Poke Out, Water Poke In, respectively, for poking in and out of the Wait port and poking in to the Reward port) with the following procedure. For each session we analyzed all correct trials. We first realigned trials to the specific event recomputing the times of occurrences of all pattern onsets with respect to the event. In each session we analyzed all transitions to patterns which occurred in at least $70\%$ of correct trials. This returned a distribution of times $\mathcal{T}(\bar{m})$ for the onset times of pattern $\bar{m}$. If the average of the distribution $\mu(\mathcal{T}(\bar{m})) \in [-0.5,0.1] sec$, we tagged the pattern $\bar{m}$ to the event. In case multiple transitions matched our criteria, we selected the one with minimum inter-quartile $iqr(\mathcal{T}(\bar{m}))$. This procedure returned patterns tagged with specific actions for each trial, cfr. \Cref{fig:4c}. We name pattern onset times $\{t_{PI}, t_{PO}, t_{WPI}\}$ respectively for the actions Poke In, Poke Out and Water Poke In.

\subsection{Decoding actions from pattern onsets}
We reversed the pattern tagging procedure to decode actions from pattern onsets. Transitions were tagged to actions using correct trials (training set) using the procedure above, then actions were decoded from pattern onsets using incorrect trials (test set). The decoding procedure follows these steps: for every trial, given an action time $t_{\bar{e}}$ and the tagged pattern onset times $\{t_{PI}, t_{PO}, t_{WPI}\}$, we classified the action according to
$$
\textrm{action} = \underset{e\in \{PI,PO,WPI\}}{\textrm{argmin}}(t_e - t_{\bar{e}}) \,\,\textrm{if}\,\, (t_{action} - t_{\bar{e}})\in [-0.5,0.1] \,\textrm{sec} \ .
$$
In case no patterns passed this criteria the action was not labelled. This procedure labelled $63\%$ of all actions. For each session and all tagged actions we estimated a confusion matrix of our decoding procedure (cfr \Cref{fig:4c}) by comparing the true actions (rows of the confusion matrix) with their predicted labels (columns of the confusion matrix). The confusion matrix across all sessions was obtained by averaging confusion matrices for individual sessions.\\

\subsection{Noise correlation analysis}
To assess trial-to-trial variability in population activity we measured the neural dimensionality of population activity fluctuations around each pattern. We first estimated the noise covariance $C_{ij}(m)$, namely, the covariance conditioned on intervals where pattern $m$ occurred (the time window with posterior probability $\ge 80\%$ in each trial):
\begin{align}
\label{methods:cm}
C(m)_{ij}= \frac{1}{N_T} \sum_a^{N_T}({r_i}^a(m) {r_j}^{a,T}(m) -  {r_i}^a(m)   {r_j}^{a,T}(m))\ ,
\end{align}
where $N_T$ is the number of trials in the session and $i,j=1,\ldots,N$ index neurons. The superscript ${}^T$ denotes vector transposition.  In each trial $a$ and window the average firing rate ${r_i}^a(m)$ in pattern $m$ was computed from Eq.~(\ref{eq:hmmrate}). We then computed the dimensionality $d(m)$ of population activity fluctuations around pattern $m$ as the participation ratio  \cite{Abbott2011,mazzucato2016stimuli}:
\begin{align}\label{methods:pr}
    d(m) = \frac{Tr[C(m)]^2}{Tr[C(m)^2]}=\frac{(\sum_i^N \lambda_i)^2}{\sum_i^N \lambda_i^2}
\end{align}
where $\lambda_i$ are the eigenvectors of the covariance matrix for $i=1,\ldots,N$ neurons \cite{Abbott2011,mazzucato2016stimuli}. This measure is bounded by the ensemble size $N$ and captures the number of directions, in neural space, across which variability is spread over.

To test the hypothesis that trial-to-trial variability is constrained within a lower dimensional subspace, we proceeded as follows. For each neural pattern $m$, we considered the first $K$ Principal Components $\{PC_1,\ldots,PC_K\}_{m}$ of $C(m)$ in Eq.~(\ref{methods:cm}), where $K$ is the integer minor or equal to the average of $PR(m)$ across the $M$ patterns within each session: $K =\textrm{floor}\left( {1\over M}\sum_{m=1}^M PR_m \right)$. This represents the across-patterns average dimensionality of noise correlations within a session. Using a Canonical Correlation Analysis we then estimated the canonical variables between $\{PC_1,\ldots,PC_K\}_{m_1}$ and $\{PC_1,\ldots,PC_K\}_{m_2}$ for pairs of patterns $m_1$ and $m_2$, obtaining the respective correlation coefficients $\rho_j$ between the $K$ canonical variables, $j\in \{1..K\}$. Alignment $A(m_1,m_2)$ was then defined as the average correlation coefficient between the canonical variables $A(m_1,m_2)={1\over K}\sum_{j=K}^N \rho_j$, cf. \Cref{fig:6b}. 


\subsection{Network model}

In this section we describe the correlated variability  model generating reliable sequences of metastable attractors (see Eq.~(\ref{model:1results})), whose dynamics is ruled by the current-based formulation of the standard rate model \cite{grossberg1969learning,miller2012mathematical}:
\begin{equation}
\label{model:1}
 \tau \dot{u}_i(t)=-u_i(t)+\sum_{j=1}^{N}J^S_{ij}\phi_j(u_j(t))+\zeta(t) \sum_{j=1}^NJ^F_{ij}\phi_j(u_j(t)) \ .
\end{equation}
The firing rates are analog positive variables given by the transformation of synaptic currents to rates by the input-output transfer function $\phi_i(u_i)$. Tranfer functions $\phi_i$ was inferred from the empirical firing rate distribution of M2 single neurons (see below section \ref{inferring_transfer_function}).
The parameter $\tau$ corresponds to the single neuron time constant. We set the M2 symmetric connectivity to be sparse \cite{mason1991synaptic,markram1997physiology,holmgren2003pyramidal,thomson2007functional,lefort2009excitatory}.
Our connectivity consists of two terms, traditionally referred to as the {\it symmetric} term $J^S_{ij}$ and the {\it asymmetric} or {\it feedforward} term $J^F_{ij}$ \cite{domany95}. The symmetric term reads
\begin{equation}
\label{model:2}
    J^S_{ij} =\frac{c_{ij}A_S}{Nc}\sum_{\mu=1}^{p} f\left[\eta^{\mu}_i\right]g\left[\eta^{\mu}_j\right] \ ,
\end{equation}
where the variable $c_{ij}$ represents the structural connectivity of the local circuit, modeled as an Erdos-Renyi graph where $c_{ij}=1$ with probability $c$. The normalization constant $Nc$ correspond to the average number of connections to a neuron; $A_S$ is the overall strength of the symmetric term.
 
The input to neuron $i$ corresponding to pattern $\mu$ was an i.i.d. standardized normal random variable, $z_i^{\mu}\overset{i.i.d.}{\sim}N(0,1)$. 

Firing rate patterns were then distributed as   $\eta_i^{\mu}\overset{i.i.d.}{\sim}{\phi(z_i^{\mu})}$, giving the symmetric term in the connectivity matrix in Eq.~(\ref{model:2}). This term represents a generalization of the covariance rule \cite{sejnowski1977storing} in which changes in the connectivity by learnig are the product of the nonlinar transformation of pre- and post- synaptic activity, i.e., $\Delta J^S_{ij}\propto f[\eta^{\mu}_i]g[\eta^{\mu}_j]$. As is shown in \cite{pereira2018attractor} this term may  lead to fixed-point attractors in this network. The functions $f$ and $g$ provide the dependence of the learning rule on the post- and pre-synaptic firing rates, respectively, and they control the firing rate statistics of the attractor. While $J_{ij}^{S}$ is symmetric only if $f=g$, we choose to keep the terminology `symmetric' for this term for consistency with early work in networks of binary neurons \cite{sompolinsky1986temporal,kleinfeld1986sequential,herz1989hebbian, domany95}. The functions $f$ and $g$ are given by the step functions 
\begin{equation}
\label{model:5}
\begin{array}{ccc}
\begin{array}{c}
f(\eta)=
\begin{cases}
q_f & \text{if } x_f\leq\eta  \\
-(1-q_f) & \text{if } \eta\leq x_f 
\end{cases}
\end{array}
&,\qquad&
\begin{array}{c}
g(\eta)=
\begin{cases}
q_g & \text{if } x_g\leq\eta  \\
-(1-q_g) & \text{if } \eta\leq x_g
\end{cases}
\end{array},\hfill
\end{array}
\end{equation}
potentiating the post- (pre-) synaptic activity with strength $q_f\in [0,1]$ ($q_g$), and depressing with strength  $1-q_f$ ($1-q_g$), respectively. The parameter $x_f$ ($x_g$) represents the threshold between potentiation and depression for the post- (pre-) synaptic dependence of the learning rule, controlling the sparseness of the nonlinar transformation of the pattern $f(\vec{\eta})$ ($g(\vec{\eta})$) imprinted in the connectivity. We assume that the average synaptic weights changes due to learning one pattern is zero, requiring $\langle g\rangle=0$, which constrains
one of the two parameters of $g$. 

The asymmetric term in Eq.~(\ref{model:1}) is the {\it correlated variability} term term $\zeta(t) \sum_{j=1}^NJ^F_{ij}\phi(u_j(t))$, where the rank $p$ of the matrix $J^F_{ij}$ is much lower than the number of neurons $N$ in the network. Hence, this term induces low-dimensional correlated fluctuations across neurons, driven by the Ornstein-Uhlenbeck process $\zeta(t)$:
\begin{equation}
    \label{model:7}
    \tau_{\zeta}\dot{\zeta}(t)=-\zeta(t)+\bar \zeta+\sqrt{2 \sigma_{\zeta}^2 \tau_{\zeta}}x(t) \ ,
\end{equation}
where $\tau_{\zeta}$, $\bar\zeta$ and $\sigma_{\zeta}^2$ are the timescale, mean and variance of the process, respectively. For a derivation of these parameters see the next section.

In Section \ref{section:private}, we compared the correlated variability model (see Eq.~(\ref{model:1})) to a {\it private noise} model
\begin{equation}
\label{model:privatenoise}
 \tau \dot{u}_i(t)=-u_i(t)+\sum_{j=1}^{N}J^S_{ij}\phi_j(u_j(t))+\bar\zeta \sum_{j=1}^NJ^F_{ij}\phi_j(u_j(t))+\sqrt{2 \sigma_p^2 \tau} \chi_i(t) \ ,
\end{equation}
where term $\sqrt{2 \sigma_p^2 \tau} \chi_i(t)$ is  additive white Gaussian noise with mean zero and variance $\sigma_p$ representing {\it private noise}, independently drawn for each neuron. Here, the asymmetric part of the synaptic couplings is constant, proportional to the parameter $\bar\zeta$, unlike the time varying asymmetric term in Eq.~(\ref{model:1}).

As a measure of the pattern retrieval (\Cref{fig:5}), we used overlaps, defined as the Pearson correlation between the instantaneous firing rate and a given stored pattern $g[\vec{\eta}^l]$ \cite{pereira2018attractor,gillett2019characteristics}
\begin{equation}
\label{model:overlap}
    m_l(t) = \frac{Cov\left[g[\vec{\eta}^l]\vec{r}(t)\right]}{\sqrt{Var(g[\vec{\eta}^l]) Var(\vec{r}(t))}} \ .
\end{equation}

\subsection{\label{methods:synaptic_noise}Two-area mesoscale model}

In this section, we show how to obtain the network model in Eq.~(\ref{model:1results}), starting from the two-area network in Eq.~(\ref{model:results1}), whose dynamics are governed by \cite{grossberg1969learning,miller2012mathematical}:
\begin{eqnarray}
\label{model:8}
 \tau \dot{u}_i(t)&=&-u_i(t)+\sum_{j=1}^{N}J^S_{ij}\phi_j(u_j(t))+\sum_{j=1}^{N_Y}W^{ M2\leftarrow Y}_{ij}r^Y_j
 \ ,\\
 \tau_{Y} \dot{r}^Y_i&=& - r^Y_i + \sum_{j=1}^NW^{Y\leftarrow M2}_{ij}\phi_j(u_j)\ ,\nonumber
\end{eqnarray}
The first equation describes the local dynamics of $N$ neurons' in area M2, with notations as in Eq.~(\ref{model:1}). The second term represents the activity $r^Y_i$ of $N_{Y}$ neurons in area Y, where we assume $N_Y\ll N$. We approximate the dynamics of the subcortical area as linear, where $W^{Y\leftarrow M2}_{ij}$ are the projections from M2 to area Y, structured by Hebbian learning (see previous paragraph) as
\begin{equation}
\label{model:9}
W^{Y\leftarrow M2}_{ij} = s_{ij}^{Y\leftarrow M2}(t)\frac{1}{N}\sum_{l=1}^py^{\mu}_i g(\eta^{\mu}_j).
\end{equation}
Here, $g(\eta^{\mu})$ is the pre-synaptic dependence of the learning rule;  $y_i^{\mu}$ is the post synaptic dependence of the learning rule, which depends on the activity in area Y. Additionally, $s_{ij}^{Y\leftarrow M2}(t)$ represents the synaptic efficacy in the Y to M2 projections due to short term plasticity \cite{tsodyks1998neural}. We consider a simplified phenomenological model capturing the temporal fluctuations in the synaptic efficacy due to STP given by 
\begin{equation}
\label{model:stp:1}
\dot{s}_{ij}^{Y\leftarrow M2}=\frac{1-s_{ij}^{Y\leftarrow M2}(t)}{\tau_{STP}}+\sqrt{\frac{2 \alpha_{Y\leftarrow M2}}{\tau_{STP}} }\xi_i^{Y\leftarrow M2}(t),
\end{equation}
where $\tau_{STP}$ corresponds to the time-scale of the fluctuations in the synaptic efficacy, $\alpha_{Y\leftarrow M2}^2$ is the variance of these fluctuations, and $\xi_i^{Y\leftarrow M2}(t)$ is a Gaussian random variable with mean zero and variance one. Here we assume that changes in the synaptic efficacy depend on post-synaptic fluctuations given by the variable $\xi_i^{Y\leftarrow M2}(t)$. We assume the activity of area Y is fast with respect to M2 ($\tau_{Y}<\tau$), replacing Eq.~(\ref{model:8}) by its steady state
\begin{equation}
\label{model:10}
    r_i^{Y} = \sum_{j=1}^NW^{Y\leftarrow M2}_{ij}\phi(u_j) \ .
\end{equation}
Feedback projections $\sum_{j=1}^{N_{Y}}W^{M2\leftarrow Y}_{ij}r_j^{Y}$ from area Y to M2 in Eq.~(\ref{model:8}) are shaped by Hebbian learning as above, obtaining
\begin{equation}
\label{model:12}
W^{M2 \leftarrow Y}_{ij} = s_{ij}^{M2\leftarrow Y}(t) \frac{1}{N_{Y}}\sum_{\mu=1}^p f(\eta^{\mu+1}_i) y^{\mu}_j \ .
\end{equation}
Similarly to the M2$\to$Y, we assume that the Y$\to$M2 projections also undergo short-term synaptic plasticity, i.e., 
\begin{equation}
\label{model:stp:2}
\dot{s}_{ij}^{M2\leftarrow Y}=\frac{1-s_{ij}^{M2\leftarrow Y}(t)}{\tau_{STP}}+\sqrt{\frac{2 \alpha_{M2\leftarrow Y}}{\tau_{STP}} }\xi_j^{M2\leftarrow Y}(t).
\end{equation}
Here, we assume that changes on synaptic efficacy of the Y$\to$M2 projections depend on pre-synaptic fluctuations given by the variable $\xi_j^{M2\leftarrow Y}(t)$. Synaptic delays in the feedback loop are long enough so that, while the pre-synaptic activity of the feedback projections correspond to the pattern $\eta^{\mu}$, the post-synaptic activity of the feedback projections is $\eta^{\mu+1}$. Then,
the input current due to the feedback loop between M2 and area Y is approximately 
\begin{eqnarray}
\sum_{l=1}^{N_{Y}} W^{Y\to{M2}}_{il} r^{Y}_l &=& \sum_{j=1}^N \sum_{l=1}^{N_{Y}} W^{{M2} \leftarrow Y}_{il} W^{Y \leftarrow M2}_{lj}\phi(u_j) \nonumber \\
&=&\frac{1}{N}\left(\bar\zeta+\frac{\sigma p}{\sqrt{N_Y}}\zeta(t)\right) \sum_{j=1}^N \sum_{\mu=1}^p f(\eta^{\mu+1}_i)g(\eta^{\mu}_j)\phi(u_j) \label{model:13}.
\end{eqnarray}
Here, we used the fact that
$\frac{1}{N_{Y}}\sum_{l=1}^{N_Y}y^{\mu}_l y_l^{\mu'}s_{il}^{M2\leftarrow Y}(t)s_{lj}^{Y\leftarrow M2}(t)$ has mean $\bar{\zeta} \delta_{\mu,\mu'}$ and finite variance $\sigma^2$, when averaged over an ensemble of $\langle \cdots \rangle_{y,\xi(t)}$ of patterns $y$ and fluctuations of the synaptic efficacies $\xi(t)$ in Eqs.~(\ref{model:stp:1}) and (\ref{model:stp:2}). Therefore, the matrix $J^F_{ij}$ in Eq.~(\ref{model:1results}) corresponds to the effective connectivity arising from the feedback loop between M2 and area Y:
\begin{equation}
\label{model:14}
J^F_{ij}=\frac{\zeta(t)}{N} \sum_{\mu=1}^p f(\eta^{\mu+1}_i)g(\eta^{\mu}_j),
\end{equation}
which has rank $p\ll N$. Assuming $p\sim \mathcal{O}(\sqrt{N_y})\ll N$, the fluctuations due to STP are order 1. We account for both the strength and the variability in the M2$\to${Y} and {Y}$\to$M2 projections, via the Ornstein-Uhlenbeck process $\zeta(t)$ in Eq. ~(\ref{model:7}). Notice that $\tau_{\zeta}$ is the effective time-scale of the temporal fluctuations in the sum over Y neurons in Eq.~(\ref{model:13}).
Its mean $\bar \zeta$ and variance $\sigma^2_{\zeta}$ control, respectively, the strength and the variability of the effective feedforward couplings obtained after integrating out the dynamics in area Y. The variance $\sigma^2_{\zeta}$ is inversely proportional to the size of the neural population in area Y. The variable  $x(t)$ represents white noise with zero mean and unit variance.

\subsection{\label{inferring_transfer_function}Inferring the transfer function from data}

For inferring the input-output transfer function 
from {\it in vivo} recordings we adapted a method
proposed in \cite{lim2015inferring} to our hidden Markov model analysis. Briefly, for each session, the empirical distribution of mean firing rates across patterns and neurons is constructed. As in \cite{lim2015inferring}, we assumed normally distributed synaptic input currents. By rank-matching the firing rates to a standardized normal distribution we obtained the empirical current-to-rate transfer function (see \ref{fig:S5}). Similarly to \cite{pereira2018attractor}, for each recorded unit we fit this curve with a sigmoidal function
\begin{equation}
    \label{transfer:1}
    \phi(u) = \frac{R_0}{1+e^{-\beta (u-h_0)}}.
\end{equation}
If input currents produced firing rates in Eq.~(\ref{transfer:1}) larger than a neuron's maximal firing rate $R_{max}$, then the correspding firing rates were set to $R_{max}$. 
Using the above procedure we inferred a distribution of parameters $\{(R_{max}^{(i)}, R_0^{(i)},\beta^{(i)}, h_0^{(i)})\}_{i=1}^{366}$, one from each recorded unit (\Cref{fig:5a}). For conveying the diversity in the transfer functions inferred from data, in our model we randomly sampled with replacement 10000 samples from the parameter distribution above.

\subsection{\label{network_simulations}Network simulations}

For the network simulations of the correlated variability model in Figs.~\ref{fig:5} and \ref{fig:6}, the parameter values used are listed in Table~\ref{table:1}. The number of sessions and trials per session are matched to those in the empirical data. The number of attractors in each session, e.g., $p$ (see Eq.~(\ref{model:2})) are taken to be the same as the number of patterns inferred in each empirical session using the HMM. An attractor was detected in the model when the overlap between network activity and the attractor is larger than 0.4. 

For the network simulations of the private noise model in Fig.~\ref{fig:S6} the parameter are the same as in Table~\ref{table:1} except $A_S = 1$, $\sigma_{\zeta}=0$, and $\sigma_p=0.2$. An attractor was detected in the model when the overlap between network activity and the attractor is larger than 0.2. 

The simulations where performed using custom Python scripts. The code is available upon request.

\begin{table}
\begin{tabular}{ |p{2cm}||p{2cm}|p{6.8cm}|  }
\hline
\multicolumn{3}{|c|}{Model parameters} \\
\hline
Parameter& Value & comment\\
\hline
$c$ & 0.1& connectivity sparsity\\
$N$ &  10000& network size\\ 
$q_f$ & .65& potentiation offset for $f$ \\
$x_f$ & 1.7&potentiation/depression threshold for $f$\\
$x_g$ & 1.7 &potentiation/depression threshold for $g$ \\
$A_S$ & 3&strength of the symmetric connectivity\\
$\bar{\zeta}$ & 0.65& mean of the synaptic noise\\
$\sigma_{\zeta}$  & 0.65&standard deviation of the synaptic noise\\
$\tau_{\zeta}$ &  20ms& synaptic noise time constant\\
$\sigma_{p}$  &0&standard deviation of the private noise \\
$\tau$ &  20ms& single neuron time constant\\
\hline
\end{tabular}
\caption{Network parameters.}
\label{table:1}
\end{table}

\acknowledgments

We would like to thank F. Cazettes, N. Steinmetz for discussions. We thank in particular L. Logiaco, J. M. Murray and N. Brunel for suggestions on the network model. This work was supported by a National Institute of Deafness and Other Communication Disorders Grant K25-DC013557 (L.M.), by the Swartz Foundation Award 66438 (L.M.).

\paragraph{Author contributions}
L.M. and Z.M. designed the project; M.M. and Z.M. performed the experiment; S.R. and L.M. analyzed the data; U.P. and S.R. developed the theoretical model with help from L.M.; L.M., S.R., and U.P. wrote the manuscript with input from M.M. and Z.M.; L.M. supervised the project.

\bibliographystyle{unsrt}
\bibliography{bib}

\end{document}